\documentclass[sigconf, nonacm]{acmart}

\usepackage[ruled,linesnumbered]{algorithm2e}
\usepackage{amssymb}
\usepackage{url}
\usepackage{diagbox}
\usepackage{svg}
\usepackage{tabularx}
\usepackage{subcaption}
\usepackage{graphicx}
\usepackage{adjustbox}
\usepackage{graphicx}
\usepackage{multirow}
\usepackage{enumitem}
\usepackage{listings}
\usepackage{hyperref}

\ifx\noeditingmarks\undefined
  \newcommand{\pgwrapper}[2]{#1\textcolor{red}{#2}}
\else
  \newcommand{\pgwrapper}[2]{}
\fi

\ifx\noeditingmarks\undefined
  \newcommand{\pgwrappergreen}[2]{#1\textcolor{green}{#2}}
\else
  \newcommand{\pgwrappergreen}[2]{}
\fi

\begin{document}
\emergencystretch 3em
% \title{Evaluating Learned Query Performance Models at LinkedIn: Challenges, Opportunities, and Findings}

\title{Evaluating Learned Query Performance Prediction Models at LinkedIn: Challenges, Opportunities, and Findings}

\author{Chujun Song}
\orcid{0009-0008-7024-7275}
\affiliation{%
\institution{University of Maryland, College Park}
    \city{College Park}
  \state{MD}
  \country{USA}
}
\email{cjsong@umd.edu}

\author{Slim Bouguerra}
\affiliation{%
  \institution{LinkedIn}
  \city{Sunnyvale}
  \state{CA}
  \country{USA}}
\email{sbouguerra@linkedin.com}

\author{Erik Krogen}
\affiliation{%
  \institution{LinkedIn}
  \city{Sunnyvale}
  \state{CA}
  \country{USA}
}
\email{ekrogen@linkedin.com}

\author{Daniel Abadi}
\affiliation{%
  \institution{University of Maryland, College Park}
  \city{College Park}
  \state{MD}
  \country{USA}
}
\email{abadi@umd.edu}

\renewcommand{\shortauthors}{Song et al.}

%%
%% The "author" command and its associated commands are used to define the authors and their affiliations.

\begin{abstract}
% Query performance prediction is a crucial prerequisite for a variety of downstream tasks like query optimization, index recommendation, system management, etc. At LinkedIn, leveraging such predictions is essential, facilitated by the availability of extensive historical query execution logs. Despite this data richness, challenges persist, such as missing critical information in historical logs, adapting models to evaluate new targets like CPU time instead of the more frequently tested latency or cardinality, new task of classification of different queries, and application on queries generated from different kinds of execution engine.

Recent advancements in learning-based query performance prediction models have demonstrated remarkable efficacy. However, these models are predominantly validated using synthetic datasets focused on cardinality or latency estimations. This paper explores the application of these models to LinkedIn's complex real-world OLAP queries executed on Trino, addressing four primary research questions: (1) How do these models perform on real-world industrial data with limited information? (2) Can these models generalize to new tasks, such as CPU time prediction and classification? (3) What additional information available from the query plan could be utilized by these models to enhance their performance? (4) What are the theoretical performance limits of these models given the available data?

To address these questions, we evaluate several models—including TLSTM, TCNN, QueryFormer, and XGBoost, against the industrial query workload at LinkedIn, and extend our analysis to CPU time regression and classification tasks. We also propose a multi-task learning approach to incorporate underutilized operator-level metrics that could enhance model understanding. Additionally, we empirically analyze the inherent upper bound that can be achieved from the models.

\end{abstract}

\maketitle

\section{Introduction}
\label{introduction}

Accurate and expedient predictions of query performance metrics, such as latency, CPU time,  memory consumption, etc, prior to the execution of queries, are critical for a host of downstream tasks, including query optimization~\cite{DBLP:journals/pvldb/MarcusNMZAKPT19, DBLP:conf/sigmod/MarcusNMTAK21}, index recommendation~\cite{DBLP:conf/sigmod/DingDM0CN19}, and query routing~\cite{DBLP:journals/pvldb/SamwelCHGVYPSTA18, DBLP:conf/sigmod/SaxenaRCLCCMKPN23}. However, traditional cost-based models are significantly limited in performing such a task~\cite{DBLP:journals/pvldb/LeisGMBK015, DBLP:journals/sigmod/JindalL22, DBLP:conf/icde/Jindal0SP21}.

Machine learning techniques are a potential alternative to using these traditional models, which employ supervised learning approaches to extrapolate insights from historical query executions~\cite{DBLP:conf/sigmod/SaxenaRCLCCMKPN23, DBLP:journals/pvldb/MarcusP19, DBLP:journals/pvldb/SunL19, DBLP:journals/pvldb/ZhaoCSM22, DBLP:conf/sigmod/Kang0TCSH21}. This approach involves the creation of a model that processes encoded representations of query plans and generates predictions. Training with historical query execution logs refines the model's capacity to accurately map queries to their corresponding performance metrics. Thus availability of rich historical execution data is crucial to the success of this approach.

% In contrast to query-\-driven methodologies, data-driven approaches often employ unsupervised learning techniques to glean insights directly from the data, thereby circumventing the necessity for pre-existing labeled training sets~\cite{DBLP:journals/pvldb/HilprechtSKMKB20, DBLP:journals/pvldb/YangLKWDCAHKS19, DBLP:journals/pvldb/YangKLLDCS20, DBLP:journals/pvldb/WangCLL21}. The core concept involves learning the joint distribution of data tuples within tables. Subsequently, the selectivity of queries is estimated by performing integration over this distribution. The crux of this approach lies in the effective utilization of a model capable of learning the intricate joint distribution of tuples.

LinkedIn's workload poses unique challenges and opportunities for the application of this approach. On the opportunity side, with over a million queries daily, there exists valuable historical data that can be used to train these models. 
%Additionally, predicting query performance before execution is crucial for downstream tasks like cost estimation and query tuning. 
On the challenges side, 
previous models are usually evaluated against synthetic benchmarks, such as TPC-H~\cite{TPCH} and JOB~\cite{DBLP:journals/pvldb/LeisGMBK015}, which fail to capture the diversity and complexity of real-world industrial workloads, raising questions about their effectiveness in real-world industrial scenarios. Additionally, LinkedIn tasks require more than just latency or cardinality regression that past work has usually focused on. For example, there are new requirements (such as classification tasks to determine the scale of the query) and new targets (such as CPU time prediction instead of latency prediction). Furthermore, historical query executions recorded in logs may lack critical query execution information. Therefore, a natural question arises: how effective will these models perform on real-world industrial workloads across a broader set of tasks where training data is more restricted? 

To address this, we assess representative models such as
TLSTM~\cite{DBLP:journals/pvldb/SunL19}, TCNN~\cite{DBLP:journals/pvldb/MarcusP19}, QueryFormer~\cite{DBLP:journals/pvldb/ZhaoCSM22}, and XGBoost~\cite{DBLP:conf/sigmod/SaxenaRCLCCMKPN23} on LinkedIn's workload. TLSTM and QueryFormer have specifically designed models suitable for tree structured query plans, XGBoost has been applied to automatic workload management in Amazon RedShift~\cite{DBLP:conf/sigmod/SaxenaRCLCCMKPN23}, and TCNN has been applied in the learned query optimizer in SCOPE~\cite{DBLP:conf/sigmod/NegiIMAKFJ21, DBLP:conf/sigmod/ZhangIM0GLFHPJ22}.

During this assessment, we find that some information, such as query structure, operator information, and plan-level metrics, are fully explored for training purposes—either as input for the model or as labels. However, other information, specifically regarding operator-level metrics, such as the number of rows and their total size, are underutilized during model training, despite their potential to enhance the model's understanding of queries and improve performance. To address this, we propose adopting a multi-task learning approach to effectively leverage this additional information. This approach is discussed in Section \ref{sec:multitask}.

% These outcomes are depicted in Figure \ref{fig:overall_comparison}. Specifically, the top left sub-figure presents a comparison based on the 90-th percentile of q-error for the latency regression task. The bottom left sub-figure illustrates the result of accuracy for the latency classification task, and the bottom right sub-figure details the inference times.

% Given the XGBoost model's exceptional performance, we extended its application to the Spark workload, considering its relevance to the query routing task that may direct queries to Spark. Detailed insights into this experiment are provided in Section \ref{sec:sparkworkload}.

We conduct experiments on two tasks: a regression task to predict the absolute metric of a query, and a query classification task.
%to categorize the queries. 
Our experiments demonstrate that an enhanced QueryFormer model, incorporating multi-task learning, achieves superior performance for regression tasks. Conversely, an XGBoost model excels in the classification task, while also achieving the fastest inference speed which is critical for real-time query routing based on prediction. For both models, CPU time prediction achieves better accuracy than latency prediction. Further insights and comprehensive analysis of these results are detailed in Section \ref{sec:experiment}.

Despite the lack of interoperability in machine learning approaches, we  investigate the upper bound on the accuracy of these models. Our investigation begins with analyzing the limitations regarding encoding the query plan, which is lossy process in which queries with distinct operator-specific information and consequently different latencies or CPU time share identical encodings. This scenario can mislead the models by presenting them with identical inputs associated with different labels. In Section \ref{sec:analysis}, we conduct a quantitative analysis to assess the impact of problems that arise due to lossy encoding.

In summary, this paper makes the following contributions:
\begin{enumerate}
    \item We conduct a comparative analysis and evaluation of various learned performance prediction models, utilizing real industrial workloads for latency regression tasks under conditions of limited information.
    \item We propose a multi-task learning approach to enhance the ability of learned performance prediction and evaluate the effect of this optimization.
    \item We evaluate the adaptability of learned performance prediction models to new tasks, including predicting CPU time, along with new
    %prediction as the new target and 
    classification tasks.
    \item We quantitatively analyze the theoretical upper bound on prediction accuracy, taking into account constraints imposed by the encoding methods used.
\end{enumerate}

\section{Overview}
\label{sec:overview}
At the time of this writing, within LinkedIn's analytical data lake infrastructure, Apache HDFS~\cite{hdfs} hosting several exa-bytes serves as the foundational data storage engine, complemented by Trino~\cite{DBLP:conf/icde/SethiTSPXSYJHSB19} and Apache Spark~\cite{DBLP:conf/nsdi/ZahariaCDDMMFSS12} as the primary query execution engines for processing analytical tasks. In this context, accurate query performance prediction is important for a variety of tasks, including cost estimation and query tuning, query routing.

In this section, we outline the tasks that could benefit from applying such techniques to LinkedIn's workloads, highlighting the associated challenges and opportunities. 

\subsection{Background}

\subsubsection{Trino at LinkedIn}
\label{sec:trino_background}
\begin{figure}[h]
    \centering
    % \begin{minipage}{0.48\linewidth}
        \centering
        \includegraphics[width=\linewidth]{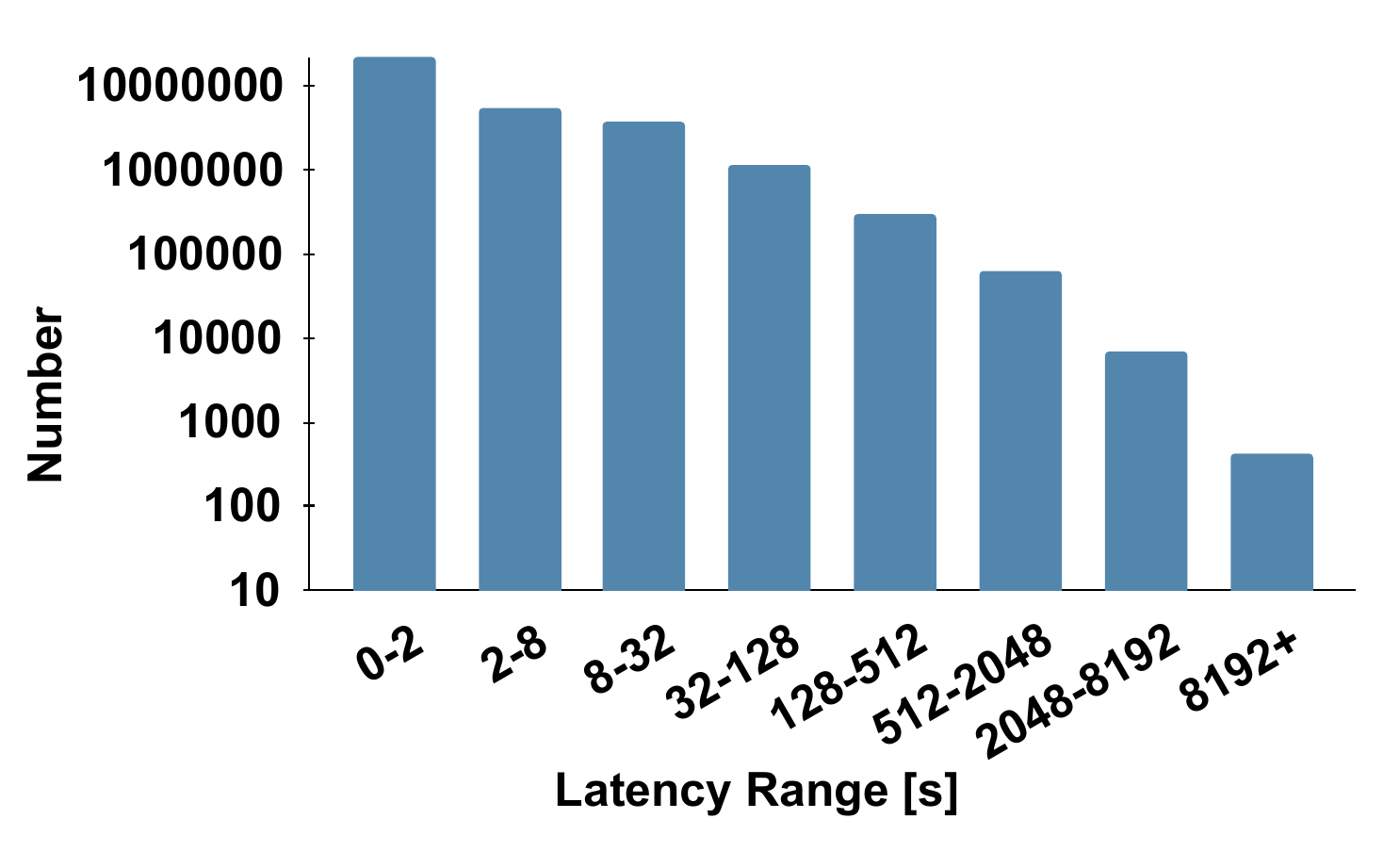}
      \caption{Trino queries latency distribution with log scale}
      \label{fig:trino_latency_distribution}
\end{figure}

Trino\cite{DBLP:conf/icde/SethiTSPXSYJHSB19} is a distributed SQL query engine designed for querying large datasets across disparate data sources. At LinkedIn, Trino is primarily employed for ad-hoc and interactive queries against data stored in HDFS. It is sometimes also used for batch data processing use cases. At the time of this writing, Trino executes over 1,000,000 queries per day with latency distribution shown in Figure ~\ref{fig:trino_latency_distribution}. From this figure, we can identify that, firstly, the range of query latencies is extensive, with smaller queries completing in seconds, while larger queries can take hours to finish. Secondly, the distribution is skewed towards shorter queries, indicating that the majority of queries are relatively quick to execute.
Given the sheer number of daily queries, it is impractical to train models on all historical data; therefore, selecting a representative subset for training is essential to manage dataset size effectively.

To meet the diverse and evolving demands of data analytics at LinkedIn—spanning interactive analytics, Extract, Transform, Load (ETL) processes, dashboarding, batch processing, exploratory analysis, and more—multiple Trino clusters have been strategically deployed. Each cluster is purpose-built and optimized to address specific workloads, ensuring high performance, scalability, and efficient resource utilization across a variety of analytical tasks.

Trino's architecture separates computation from storage. It implements the computation engine but delegates storage system management to external systems. 
Since it does not directly control storage, essential statistics such as maximum, minimum, and other aggregate data about data in different table columns (end even sometimes basic information such as number of rows in a table) are often unavailable. This gap leaves Trino reliant on heuristic-based rules for query optimization instead of more traditional cost-based optimization approaches.

% \subsubsection{Spark at LinkedIn}

% Similar to Trino, Spark also implements a computation engine for generic data processing. However, unlike Trino, whose primary interface is SQL, Spark supports multiple different intefaces and data manipulation languages so that it can be used in a wider spectrum of use cases in areas of data engineering, data science, and machine learning. At LinkedIn, Spark is used for a variety of data processing needs including data management, data pipelines, ad-hoc queries, and various machine learning related tasks. In this paper, we focus on Spark's relational query execution capabilities, including queries written in SQL and Spark's declarative DataFrame APIs, both of which are widely used at LinkedIn.

% Since Spark's architecture also decouples storage  from the computation engine, it suffers from the same issues that Trino runs into described above. At LinkedIn, Spark applications are executed on Apache Hadoop YARN or Kubernetes and resources are acquired on-demand on a per-application basis. This enables a highly scalable and resilient platform, with robust isolation between applications, but often incurs high startup overheads, such overhead can be in the order of minutes. 

\subsubsection{Uses for Query Performance Prediction}
\label{sec:usage}

The ability to predict latency and resource consumption characteristics of queries with reasonable accuracy can be used to solve many important problems in LinkedIn's data processing ecosystem. Two such examples are shared below. %, with the primary focus of decreasing both the cost of query execution as well as the amount of user input required.

\noindent\textbf{Cost Estimation}
A common issue faced throughout the industry is the lack of visibility into the cost of a query, \textit{before} it is submitted, where we define the cost as the hardware resource consumption of the query. 
%It is typical that a user who is about to submit a query has no idea how long it will take to execute
Users often have no independent ability to estimate this cost prior to query execution, only learning through trial and error as they run similar queries on similar datasets. 
%Similarly, users may not be aware of the cost of the query, and
Even if this information is exposed to a user, it is typically only made available after the query has been executed, at which point the cost has already been incurred. Accurate estimates of cost and latency can help users make informed decisions about whether they want to submit a query, and help them quickly notice mistakes that may lead to very expensive queries, such as forgetting to add a partition filter.

\noindent\textbf{Query Tuning}
% Spark has an enormous array of configuration parameters that can be used to optimize its execution behavior, and the deployment model for Spark at LinkedIn requires users to declare the resources required by their application in terms of number of CPU cores per executor, memory per executor, and executor count. While some dynamism exists in the system (such as Dynamic Resource Allocation), this is primarily a manual process carried out by users, often using trial and error.
Trino offers a range of tuning-related properties\footnote{https://trino.io/docs/current/admin/tuning.html}, such as those controlling memory allocation, parallelism, and scheduling, enable users to address the unique demands of diverse query types. Additionally, it provides administrators with significant flexibility to optimize clusters for specific workloads. Accurate query performance prediction can not only help improve the accuracy and effectiveness of tunning, but also allowing the parts of the optimization space to be explored in an automated fashion, reducing the burden on these administrators who would otherwise perform many of these optimizations manually.

\noindent\textbf{Query Routing}
Different Trino clusters are deployed to meet diverse requirements for query analytics, necessitating the routing of queries to the most suitable cluster based on their characteristics. For some queries, determining the appropriate cluster heuristically can be challenging. To address this, learned models can be leveraged to analyze query characteristics in advance, enabling more informed and accurate routing decisions.

% By integrating runtime information into the features used for prediction, Trino can estimate query cost and latency across various runtime environments. This allows for automated decision-making without requiring user intervention.

% Example uses include making Spark resource request decisions without user input routing Trino queries to different clusters optimized for different query shapes, and even routing queries \textit{across} engines such as Trino and Spark depending on which engine is better suited for the query.

% \noindent\textbf{Query Tuning}
% Example uses include making Spark resource request decisions without user input routing Trino queries to different clusters optimized for different query shapes, and even routing queries \textit{across} engines such as Trino and Spark depending on which engine is better suited for the query.

\subsection{Challenges}
The industrial context of the workload mentioned above introduces several challenges for the application of predictive models. 

\noindent \textbf{The complexity of large number of real-world queries:}
%On the one hand, the sheer volume of queries is already substantial, with around one million queries daily for Trino, making it 
We mentioned above that at one million daily queries at LinkedIn, it is impractical to train on the entire set of historical queries. 
Additionally, these queries access tens of thousands of tables, posing significant challenges for table encoding, since the commonly used method of one-hot encoding cannot scale to this level. Moreover, the scale of the queries, in terms of latency or CPU time, spans a broad spectrum, ranging from sub-second short interactive queries to long ETL queries that can take days to complete. The queries themselves can be highly complex, with some containing thousands of operators and accessing multiple terabytes of data across dozens of tables. Furthermore, these systems are typically deployed within large-scale multi-tenant clusters, leading to inconsistent query performance even for the same exact query.

\noindent \textbf{The unavailability of crucial information:}
\label{sec:infounavailable}
%LinkedIn's extensive utilization of Trino, which separate computation from storage, poses unique challenges. This separation prevents these engines from directly managing the storage system, often resulting in incomplete table statistics. 
As described above, LinkedIn's use of virtualization software such as Trino often leads to an absence of comprehensive data statistics.  This complicates the task of encoding complex table characteristics, which is an integral part of the application of such methods~\cite{DBLP:journals/pvldb/SunL19, DBLP:journals/pvldb/ZhaoCSM22}. 
Moreover, during the process of query planning and optimization, some internal symbols specific to a query plan may be generated. Capturing the derivation of these symbols can be challenging, making predicates or conditions containing them difficult to parse.

Listing~\ref{lst:sqlquery} illustrates this issue. The upper section of the listing shows a query snippet from TPC-H Q16, while the lower section displays the corresponding part of the execution plan generated, with irrelevant information removed. The internal symbol \texttt{expr} in the plan is a boolean derived from the SemiJoin, which checks if the \texttt{ps\_suppkey} column is equal to the \texttt{s\_suppkey} column. However, such information is not explicitly presented in the plan, and extracting this information in a systematic manner is difficult.

\begin{lstlisting}[caption=SQL query and plan snippet, label=lst:sqlquery]
ps_suppkey NOT IN (
    SELECT s_suppkey
    FROM
      postgresql_1gb.public.supplier
    WHERE
      s_comment LIKE '%Customer%Complaints%'
)
-----------------------------------------------
FilterProject[filterPredicate = (NOT "expr")]
  Project[]
    SemiJoin[ps_suppkey = s_suppkey]
\end{lstlisting}

% Moreover, while LinkedIn's repository of past execution logs is voluminous, these logs do not provide a complete picture of the query execution plan. Certain predicates and conditions within these logs incorporate symbols unique to individual queries, and the logs often omit the derivation of these symbols from the original columns. This omission presents a significant barrier to learning a consistent interpretation of these symbols across different queries, complicating the application of query-driven approaches in accurately predicting query performance within LinkedIn's dynamic and complex analytical landscape.

\noindent \textbf{The necessity to adapt models for a broader range of tasks:}
Section ~\ref{sec:usage} discussed some applications of traditional latency and cardinality regression tasks at LinkedIn along with some new tasks. In addition, LinkedIn needs to implement models to facilitate query classification. For example, based on the classification of a query, e.g. whether it is a "small query" or a "large query", the SQL Gateway router may route a query to a particular query execution engine. In this context, cardinality prediction becomes less relevant, as the primary concern shifts to query-level metrics that differ across query engines, such as CPU time usage.

\subsection{Research Questions}
\label{sec:research_question}
These challenges pose unique challenges to the application of performance models to real-world industrial workloads. We thus investigate the performance of such models in this context, which we break down into the following four questions:

\begin{enumerate}[label={}, leftmargin=*]
    \item \textbf{Q1:} How do these models perform on real-world industrial data with limited information available (recorded) in the log?
    \item \textbf{Q2:} Can these models generalize to new tasks, such as CPU time prediction and classifications?
    \item \textbf{Q3:} Can other information available in this setting be used by such models to improve their performance?
    \item \textbf{Q4:} What are the theoretical performance limits of these models given the available data?
\end{enumerate}

To address research questions 1 and 2, we assess several representative techniques:TLSTM~\cite{DBLP:journals/pvldb/SunL19}, TCNN~\cite{DBLP:conf/sigmod/MarcusNMTAK21}, XGBoost~\cite{DBLP:conf/sigmod/SaxenaRCLCCMKPN23}, and QueryFormer~\cite{DBLP:journals/pvldb/ZhaoCSM22}, and investage how well they perform on LinkedIn workloads for various tasks.
The results of this evaluation is reported in Section~\ref{sec:models_comparison}. For research question 3, %we analyzed the models and found 
we find that operator-level metrics are often overlooked, and propose the use of multi-task learning to better utilize these metrics, detailed in Section~\ref{sec:multitask} with results in Section~\ref{sec:multi_task_res}. For research question 4, we analyzed the theoretical upper bound on the performance achievable by these models, with our analysis detailed in Section~\ref{sec:analysis}. 

\begin{figure*}[htbp]
    \centering 
    \includegraphics[width=0.9\textwidth]{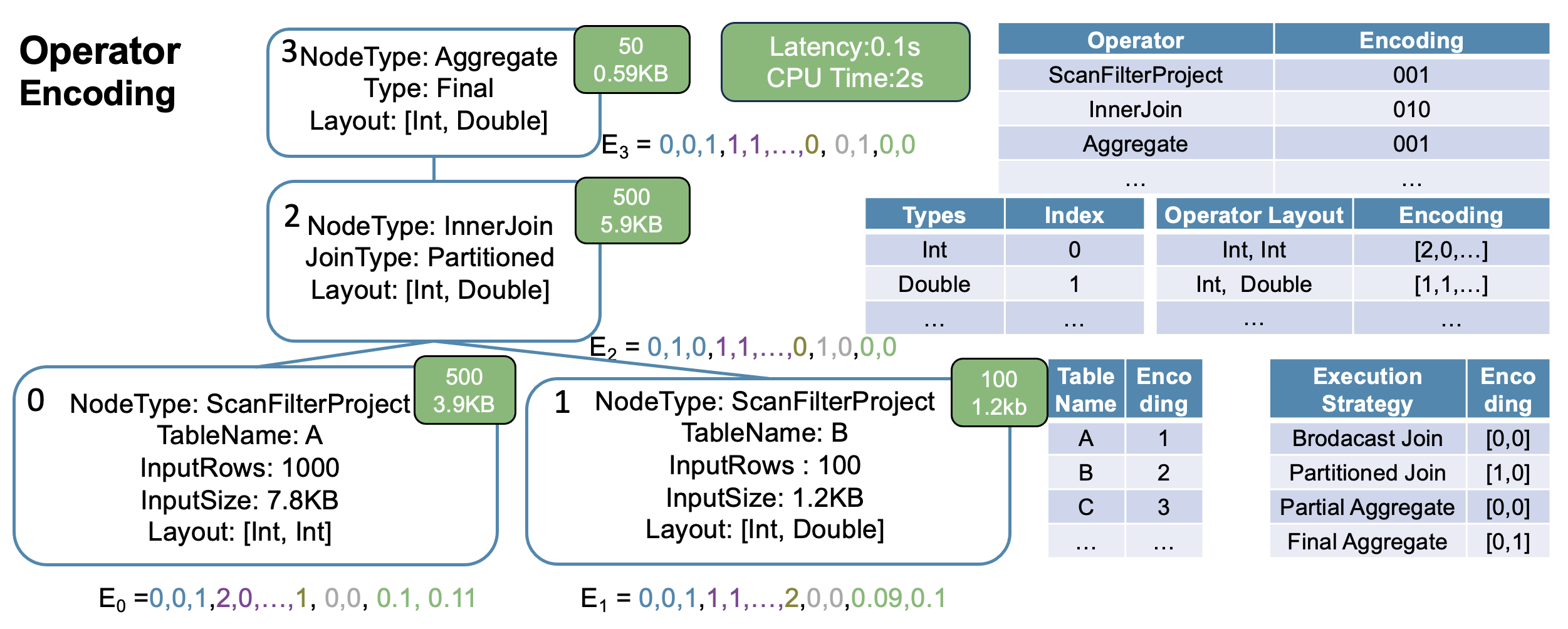}
    \caption{Deep learning models operator-level encoding example}
    \label{fig:nn_encoding}
\end{figure*}

\section{Design}

In this section, we first describe the data available in the query execution log (Section ~\ref{sec:queryexecutionlog}), and then (in Section~\ref{label:encoding}) describe the methods employed for encoding this log for neural network based models (TLSTM, TCNN, and QueryFormer) and a machine learning model: XGBoost. Subsequently (in Section ~\ref{sec:multitask}), we detail the approach of leveraging multi-task learning to utilize operator-level metrics effectively. Finally (in Section ~\ref{sec:classification_task}), we detail our strategy for performing the classification task, with a particular emphasis on class assignment. 

\label{sec:design}

\subsection{Query Execution Log}
\label{sec:queryexecutionlog}
The data available in the query execution log consists of SQL query execution plans annotated with runtime statistics both at the operator level and plan level. For operator-level statistics, it contains information such as the cardinality and size of the outputs of specific operators. For plan-level statistics it contains the latency and CPU time of the query, where latency is defined as the wall-clock end-to-end execution of a query, and CPU time is the sum of CPU time consumed by each individual node that participates in executing the query. The left side of Figure~\ref{fig:nn_encoding} gives an example of such a query execution log.

\subsection{Query Encoding}
\label{label:encoding}

Both machine learning and deep learning models require numerical values as inputs, yet query plans are represented as trees with operator information encoded in each node. Therefore, an initial feature extraction step is needed to transform these structures into a numerical format compatible with these models. Since machine learning and deep learning models typically require different forms of input, we will describe the encoding strategies for each model type separately. Additionally, we %address the challenges associated with encoding brought because of the challenge of the 
discuss the encoding difficulty challenge arising from  the large number of tables, as overviewed in Section \ref{introduction}.

\subsubsection{Neural Networks}
\label{sec:nnencoding}
%For the encoding part, because the 
SQL query plans are represented by a tree, with operator information in each node. Since the operator information is hetergeneous, 
%in the tree has heterogeneous characteristics for the operator information 
as illustrated in the left side of Figure \ref{fig:nn_encoding}, we utilize a two-level encoding approach to capture the overall structural information and operator-level information. The operator-level encoding is the same for all neural network models, while the overall structural encoding is where the models differ.

For operator-level encoding, some information is common across all operators, %each operator would convey various parts of information, some are common across operators, like 
such as the category of the operator, the columns they produced, 
%by these operators 
and their data type information (we refer to this typing information as operator layout). Other information is specific to individual operators, such as the one or more input tables, and statistics about these inputs. %from the scan operator. 
The heterogeneous nature of this operator information makes encoding it in a uniform manner difficult. %To address this, we 
We therefore initially use a uniform encoding only for the information which is consistent across operators, 
%delineate distinct parts within an operator. Common parts are encoded uniformly across all operators, whereas parts 
while information unique to specific operators are encoded exclusively for those operators, leaving it blank for operators for which this information is inapplicable. Ultimately, we amalgamate the encoded segments to formulate a comprehensive operator encoding.

%The parts we identify are operator types, operator layout, table names, operator execution strategy, and input data statistics, where operator types and operator layout are common, and table names, operator execution strategy, and input data statistics are specific for some operators. 
When encoding operator type $E_t$, we use one-hot encoding, since the cardinality of operator types is limited. For operator layout encoding $E_l$, we use a fixed-sized vector with a length equal to the number of data types in the type system. Each position in the vector represents a specific type, while the value in that position indicates the number of times this type appears in the layout. For example, if there were six types: Integer, Double, String, Date, Map, and Array, we would use an encoding template of \{0,0,0,0,0,0\}, where the first position represents the number of Integers, and the second position represents the number of Doubles, etc. If the operator layout was [Integer, Integer, Double, Map]; then the encoding would be \{2,1,0,0,1,0\}, meaning Integer appears twice, while Double and Map appear once. 

Encoding table names $E_{tn}$, has traditionally been done via one-hot encoding, since table cardinality is usually limited. However, in industrial workloads, there can be hundreds of thousands of tables\footnote{Many of these tables are temporary tables generated within queries.}, to which one-hot encoding cannot scale. Instead, we use dictionary encoding. %We can obtain the names of all tables by iterating all data and then assigning a unique id to each table name.

For operator execution strategy $E_s$ (e.g broadcast join vs. partitioned join, or whether an aggregate is partial or final), we use a fixed-size indicator vector, where the vector length is the total number of execution strategies, with each position referring to a specific execution strategy. For example, to encode the execution strategies for joins and aggregates, the length of the vector would be 2, where the first position encodes whether it is a partitioned or broadcast join and the second position encodes whether the aggregate is partial or final. A partitioned join operator would thus be encoded as \{1,0\}.   

For table input statistics $E_{is}$, due to the lack of detailed statistics available in the log, as explained in Section \ref{sec:trino_background}, we only encode the number of input tuples and their overall size. Since these numbers, like data size, could span a large range from several bytes to terabytes, we leverage min-max log normalization to normalize these values. We first find the minimum value (min) and maximum value (max) of the category by iterating through all training data, and then use the following formula to normalize the encoded value:
\begin{equation}
Norm(x) = \frac{\log(x) - \log(\min)}{\log(\max) - \log(\min)}
\label{eq:minmaxlog}
\end{equation}
%To deal with the heterogeneous nature of operator-specific information (for example, a non-leaf operator would not have input statistics information), we choose to set these vectors to zero for values that don't have such information. With that, the 

The final encoding of a node $E_n$ can thus be represented as a concatenation of all these vectors:
\begin{equation}
E_n = Linear(E_t)||Linear(E_l)||Linear(E_{tn})||Linear(E_{s}) ||  Linear(E_{is})
\end{equation}, where "Linear" is a specific linear layer designed for each value. Zeros are placed in this final encoding for values that are not applicable (for example, a non-leaf operator would not have input statistics information).

Figure \ref{fig:nn_encoding} illustrates this encoding method. % discussed previously. 
The left side of the figure shows the input query plan along with detailed operator-level information, including metrics. The operator-level metrics, positioned in the upper-right corner of each operator's rectangle, will be integral for the multi-task learning approach outlined in Section \ref{sec:multitask}. 
%Central to the figure are the overall query execution latency and CPU time metrics for the query. 
The right side of the figure shows the encoding techniques we discussed above: one-hot encoding for operator types, dictionary encoding for table names, and fixed-size vector encoding for operator layouts and operator execution strategy. The encoding of input statistics is not shown.
%, calculated as per the formula in Equation \ref{eq:minmaxlog}, is not depicted in the visual representation. 
Near each of the four operators, their final encodings are displayed, with varying colors used to differentiate among the distinct encoding components. 

Similar to the work on AutoSteer~\cite{DBLP:journals/pvldb/AnneserTCXPLM23}, we do not encode predicates and conditions,  
%are not encoded due to the following rationale: our dataset comprises only past query plans, within which predicates and conditions 
since they may contain symbols (e.g., expr\_1, expr\_2, etc.) that are uniquely meaningful within the confines of a single plan, 
%. Sometimes, the column names of such symbols from which they are derived—
and for which their column derivations are not documented as we have shown in Section ~\ref{sec:infounavailable}.
%, rendering us unable to trace their derivation, making learning a consistent meaning of predicates and conditions across plans difficult. Additionally, it's worth noting that AutoSteer~\cite{DBLP:journals/pvldb/AnneserTCXPLM23}, which similarly processes Presto\cite{presto} \footnote{The current code bases for Presto and Trino share a common ancestor, also known as Presto. The code bases have since diverged but the fundamental architecture remains similar, thus research findings are generally transferable across the two systems.} input, adopts an identical methodology.

\begin{figure*}[htbp]
    \centering 
    \includegraphics[width=\textwidth]{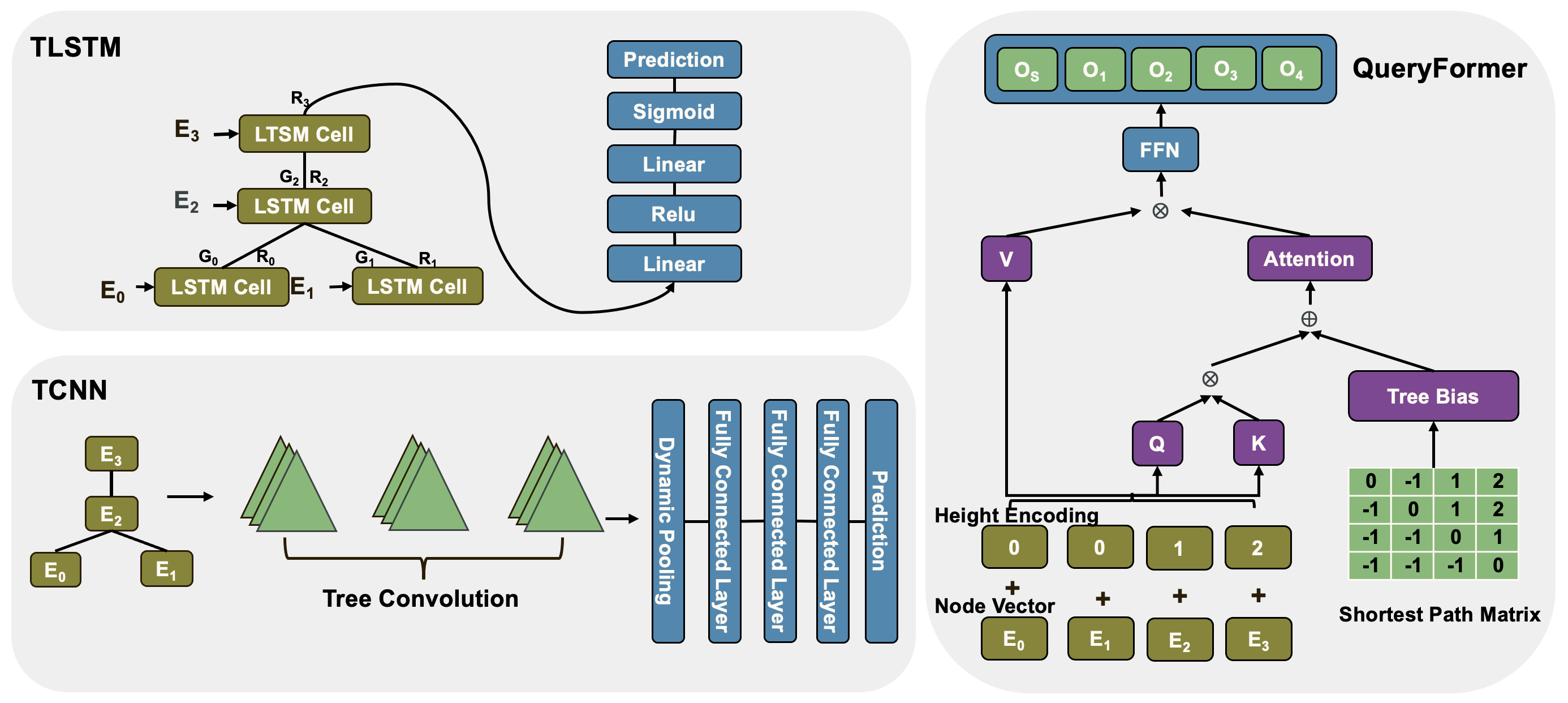}
    \caption{Deep learning models plan-level encoding example}
    \label{fig:models}
\end{figure*}

For encoding the structural information of query plans, various neural network models employ distinct, tailored architectures as depicted in Figure \ref{fig:models}. The operator encodings, represented by vectors $E_{i}$, are consistent with those in Figure \ref{fig:nn_encoding}.

The TLSTM model utilizes a tree-based LSTM structure that mirrors the query plan's topology, as shown in the upper-left subsection of Figure \ref{fig:models}. Each LSTM cell corresponds to a specific node in the query plan, receiving operator-level encodings as inputs. The information flow between LSTM cells symbolizes the connections between different operators, and are controlled by vectors [$G_{i}$, $R_{i}$]. The final output ($R_{3}$ in the figure), which represents the overall query encoding, is then processed by an estimation network for prediction.

TCNN employs a tree-structured convolutional neural network to represent the query plan's structural details, as illustrated in the lower-left subsection of Figure \ref{fig:models}. The approach begins by converting the query plan into a fully binary tree, where each node has either zero or two children. A triangular convolutional kernel traverses the tree, capturing the relationship between a node and its immediate children. Subsequent to tree processing, the distilled structural information is forwarded to an estimation layer for prediction.

QueryFormer adapts the transformer architecture by making two modifications to accommodate the tree structure. First, it substitutes positional encoding with height encoding, thereby encapsulating the depth of nodes within the query plan. It integrates the length of the pairwise shortest path between nodes, termed tree bias, into the original attention matrix. This ensures parental nodes exclusively "attend" to their descendants. The resulting attention scores are directed into a dense neural network for final prediction, outputting a 
$N  \times D$ matrix where each row denotes the processed representation of a node, as illustrated by $O_i$s in the figure. To aggregate a holistic query plan representation, QueryFormer introduces a "super node", akin to the [CLS] token in BERT~\cite{DBLP:conf/naacl/DevlinCLT19}, resulting in a 
$(N+1) \times D$ output matrix. The super node's representation is fed into the estimation layer for final prediction. %For detailed descriptions, readers are encouraged to read the respective papers.

%One distinction between our methodology and those previously described lies in the handling of operators with more than two children. Existing methods typically 

Each of these methods limit each operator
%presuppose that each operator is limited
to a maximum of two child operators. However, there are several operators in Trino that have more than two children, 
%queries, certain operators, 
such as exchange operators which can have dozens of children.
%, can have significantly more children, sometimes extending into the dozens. 
%To address this issue, we consider two potential solutions: (1) adapting the network to accommodate n-ary operators directly, or (2) modify the plans to decompose n-ary operators into a series of binary operators, a strategy recommended by BAO~\cite{DBLP:conf/sigmod/MarcusNMTAK21}. Given the relative rarity of n-ary operators—with the majority of operators being unary or binary—we opt for the latter approach. This decision is further validated by experimenting with modifying the TLSTM structure to incorporate n-ary operators directly, which yields inferior results compared to the binary operator splitting method.
To address this issue, we use the strategy recommended by BAO~\cite{DBLP:conf/sigmod/MarcusNMTAK21} to modify the plans to decompose n-ary operators into a series of binary operators.

For encoding the label of the regression task, given the significant variance in latency (ranging from seconds to one day) and CPU time (from seconds to several days), we employ min-max log normalization, as described in Equation \ref{eq:minmaxlog}, to smooth this variance.

\subsubsection{XGBoost}
\begin{figure}[h]
        \centering
        \includegraphics[width=\linewidth]{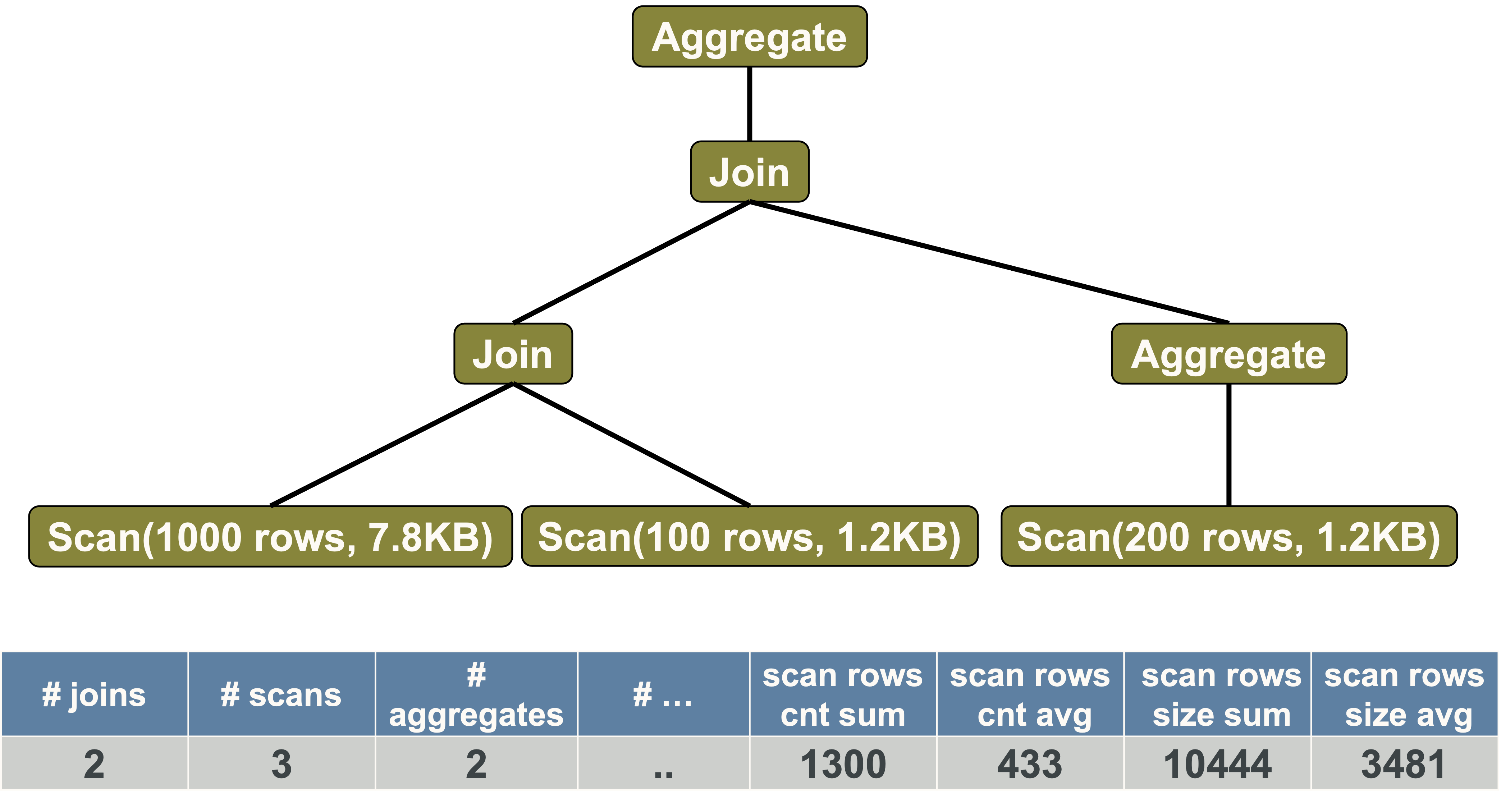}
        \caption{XGBoost encoding example}
        \label{fig:xgboost_encoding}
        \Description{Number of Spark queries per day across a month}
\end{figure}

Due to XGBoost's limitation in processing only numerical and categorical data, without the capability to encode more complex data structures, we adopt a more direct encoding approach as outlined in Saxena et al.~\cite{DBLP:conf/sigmod/SaxenaRCLCCMKPN23}. This strategy involves encoding easily quantifiable information such as the frequency of different operators within the plan tree, the total cardinality of input rows, and their cumulative size across all scan operators. Additionally, we calculate and encode the average values of these metrics for all scan operators. The frequency of different operators within the plan tree can be calculated by iterating over the plan tree and counting different operators. The total cardinality and size of all input rows can be obtained from the table statistics. Figure \ref{fig:xgboost_encoding} depicts the encoding process for an example plan. The resulting encoded representation of the plan is [2, 3, 2, ..., 1300, 433, 10444, 3481]. 

% In the case where there are operations that are pushed into the underlying system, we can rely on the 

% Figure \ref{fig:xgboost_encoding} depicts the encoding process for an example plan. The resulting encoded representation of the plan is [2, 3, 2, ..., 1300, 433, 10444, 3481]. This encoding procedure is applied to every plan within the training dataset, and the XGBoost model is subsequently trained using these encoded representations.

Given that boosting trees are largely insensitive to input scale, input data normalization is not required. However, for labels in regression tasks, we employ log normalization as follows:
\begin{equation}
Norm(x) = \log(x)
\end{equation}
We found this approach yielded superior results relative to min-max log normalization. % is also tested, log normalization yields superior results and is therefore preferred.

% The encoding of Spark plans follows a similar pattern, except we also concatenate the values for \texttt{num-executors}, \texttt{executor-cores}, and \texttt{executor-memory} into the encoding to take into account the cluster size of Spark into consideration.

\subsection{Multi-task Learning for QueryFormer}
\label{sec:multitask}
The QueryFormer\footnote{We are unable to perform a similar process for TCNN or XGBoost, as these methods do not provide operator-level outputs.} 
%architecture predominantly concentrates on the role of attention between different operators in influencing overall query performance; however, it 
approach does not fully capitalize on operator-level metrics, such as the cardinality of rows produced by the operators, which are accessible during training and are highly correlated with the execution plan's overall performance. To incorporate such operator-level metrics, we employ multi-task learning~\cite{caruana1997multitask}, to not only estimate the overall performance of the execution plan but also to predict the specific metrics of the operators within the plan. This dual focus aids in better capturing the characteristics of operators, potentially enhancing the accuracy of query performance predictions.

\begin{figure}[h]
        \centering
        \includegraphics[width=0.9\linewidth]{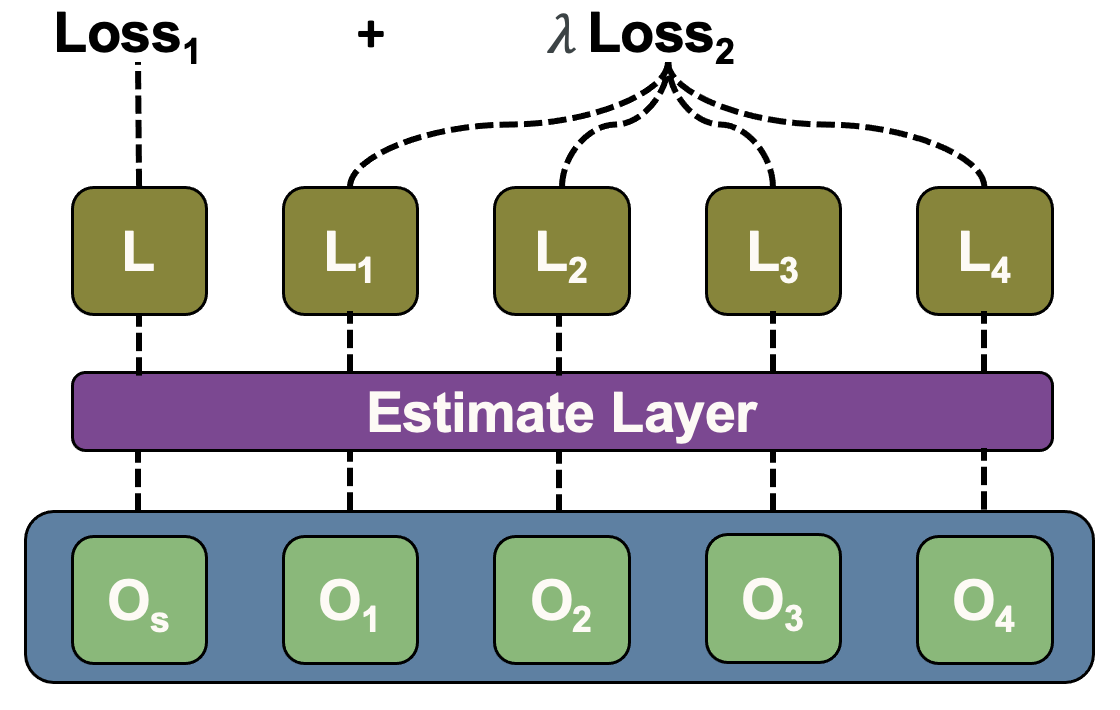}
        \caption{Multi-task learning example}
        \label{fig:multi_task_learning}
        \Description{Number of Spark queries per day across a month}
\end{figure}

To integrate the task of learning operator-level metrics, it is necessary to establish a fine-grained representation for individual operators. The output node embeddings $O_i$ in the QueryFormer architecture, which correspond to the encoding of input nodes $E_i$,  effectively represent different operators. We introduce an additional estimation layer that utilizes these embeddings to calculate specific operator-level estimates, and then try to minimize the loss in both operator-level metrics and overall plan metrics at the same time. This approach is depicted in Figure \ref{fig:multi_task_learning}, where the $O_i$s correspond to the output of the QueryFormer model, $L$ correspond to the plan level label, and $L_i$ correspond to the operator level label, as shown in Figure \ref{fig:nn_encoding}. The final $Loss$ is defined as \begin{equation}
    Loss = Loss_1 + \lambda Loss_2
\end{equation}  where $Loss_1$ represents the loss at the plan level, $Loss_2$ denotes the operator level loss, and $\lambda$ is the hyper parameter. We opt for a linear combination of these losses, given their conceptual similarity, to formulate the final loss metric. We experimentally evaluate this optimization strategy in Section \ref{sec:models_comparison}.

\subsection{Classification Task}
\label{sec:classification_task}
For many applications, generating a precise prediction of latency or CPU time is not as important as generating a high level description of a particular query plan; %critical as understanding the query's magnitude in terms of these metrics—for instance, 
for example, detecting whether a query is large or small in order to route different queries to different clusters. 
%This perspective shifts the focus toward classification rather than regression, aiming to ascertain the category to which a query's latency or CPU time belongs. 
This is essentially a classification task instead of a regression task. 
When approaching this goal as a classification task, the queries' labels are transformed from absolute numbers into classes based on a predefined classification function. This involves establishing class boundaries and then assigning queries to these classes according to their latency or CPU time. For example, if the class boundaries are set at 1 second, 1 minute, 10 minutes, and 1 hour, queries with latencies under 1 second would be classified into class 0, those with latencies equal to or greater than 1 second but less than 1 minute into class 1, and so forth.

During model training, these class labels are employed as targets, necessitating an adjustment in the model's output to predict a class rather than a numerical value. To optimize the model for classification, cross-entropy loss is utilized as the training criterion.

The determination of the number of classes is based upon the task's requirements. A binary classification might suffice for distinguishing between queries of short and long durations. However, for more nuanced distinctions, a greater number of classes may be needed. Our experiments in Section \ref{sec:num_class} find that 
%It's important to note that 
the chosen number of classes can significantly impact classification accuracy.
%, detailed in Section .

\section{Implementation}

The implementation of the existing learned performance prediction models have been extensively discussed in their respective papers. In this section, we discuss some problems specifically related to dataset preparation for the industrial workload. 

\subsection{Data Generation}
\label{sec:datagen}
Training data for machine learning models can be obtained from logs of historical executions. 
%Considering the large number of query execution plans, direct training on the entire dataset is impractical. As a workaround, we opt for 
We randomly sample queries from these logs to create our training set. However, 
%this method faces a challenge: the predominance of short-duration queries in Trino query logs. As 
as illustrated in Figure \ref{fig:trino_latency_distribution}, the distribution of query latencies in the Trino query logs reveals a skew towards shorter queries. %Random sampling, therefore, risks biasing the training dataset towards these shorter queries, potentially leading to a model that underestimates the latency of longer queries. To counteract this, we 
We therefore implement biased sampling to ensure a higher representation of longer queries in our training dataset.

\begin{table}[h]
\centering
\caption{Trino dataset characteristics}
\begin{tabularx}{\linewidth}{@{} *{4}{>{\centering\arraybackslash}X} @{}}
\toprule
\textbf{Dataset} & \textbf{Number} & \textbf{Distinct Shapes} \\
\midrule
Training & 359870 & 12738\\
Testing & 35050 & 3531 \\
% spark\_training & 2221690 & 9244\\
% spark\_testing & 69248 & 3044\\
\bottomrule
\end{tabularx}
\label{tab:trino_spark_dataset}
\end{table}

\subsection{Data Cleaning}
\label{sec:dataclean}
Unlike queries derived from synthetic datasets such as TPC-H or IMDB, queries from real datasets typically contain noise, which can make their performance incompatible with the query plan and input statistics. 
%Taking Trino queries as an example, some may have a 
For example, some may have a simple structure and scan minimal data, yet exhibit prolonged latency. This could be due to various reasons: results might be buffered in the output operator and fetched slowly by the upstream client; there might be an extended shuffle time due to network issues; or there could be unexpected delays in response times from external systems. 
%Moreover, some queries may only access system tables, which we are not interested in, and could constitute a significant percentage of the overall query volume. The identification of such queries generally relies on domain knowledge, necessitating their exclusion from the training data for accuracy.

%To address these issues, we aim to detect and eliminate problematic queries from the dataset. Specifically, if a query accesses any system table, it is removed. Additionally, 
Therefore, queries were eliminated from the training dataset if there was a significant disparity between the total amount of data scanned and the end-to-end latency based on a defined threshold ratio.
%between end-to-end latency and the amount of data scanned; queries exceeding this threshold are wiped out. 
Queries from specific clients can have a prolonged execution time, and we exclude such queries from our training dataset.
Furthermore, queries that exclusively access system tables were removed since they were not of interest to our study.

% \subsection{Training}
% \subsubsection{Neural Networks}
% Deep learning training is inherently time-intensive, to expedite this process, we have devised strategies to accelerate training. The overall process is as follows: CPU reads the training data from the disk, pre-processes it, and subsequently feed the processed data to the GPU for neural network training. For the data reading component, we leveraged the multi-process data loading feature provided by PyTorch's data loader to enable parallel data loading. For data preprocessing, we conducted code profiling to identify and optimize the most computationally expensive functions. For instance, in the training of QueryFormer, preprocessing necessitates calculating the pairwise shortest path between all nodes in the query plan. This step represents a significant bottleneck, which we addressed by employing the Floyd-Warshall algorithm for optimization. Regarding the neural network training segment, we adopted distributed training across four GPUs utilizing HuggingFace's Accelerate\cite{accelerate} framework. 

% \subsubsection{XGBoost}
% Training on XGBoost is fast and can be done on local machine. Section \ref{sec:timecompare} compares the training time of different models. We use exhaustive search to find the optimal configuration of parameters including \textit{eta}, \textit{gamma}, \textit{max\_depth}, \textit{min\_child\_weight} etc.

\section{Experiment}
\label{sec:experiment}
In this section, we present the results of an experimental study that was designed to shed light on the research questions described in Section~\ref{sec:research_question}. We compare different models for latency regression, and explore the generalizability of the models to classification tasks, and CPU time prediction as a new target. Additionally, we examine the effects of multi-task learning, number of classes in classification tasks, and larger datasets.

% All the experiments are conducted on the Trino workload, except for the results in Section~\ref{sec:sparkworkload}
% , where we present the results on the Spark workload using XGBoost.
\subsection{Set Up}
\label{sec:expsetup}
\noindent \textbf{Dataset} Table \ref{tab:trino_spark_dataset} lists the characteristics of the datasets. Trino datasets are obtained by utilizing the biased sampling method discussed in Section \ref{sec:datagen} and the data cleaning approach discussed in Section \ref{sec:dataclean} from one week of real query execution logs. 

% and Spark datasets are obtained in a similar fashion from a month's volume of Spark queries. 

The shape is defined as the overall structure of the plan, with other information such as operator level information omitted. 

\noindent \textbf{Training}
The neural networks are trained on 4 A100s with 40 GB memory each. We utilize HuggingFace Accelerate\cite{accelerate} for distributed training. We train the model until the monitored metrics didn't increase for 3 epochs. XGBoost is trained on a MacBook Pro equipped with Intel(R) Core(TM) i9-9980HK CPU @ 2.40GHz and 32 GB RAM. For the stop condition, we set the parameter \textit{early\_stopping\_rounds} to 10. 
%We fine-tune all the models for better accuracy.

% The reason we have more training samples for the Spark side is that Trino needs to be trained on deep learning models which is far more time-consuming compared to the training of the XGBoost model for Spark. Section \ref{sec:models_comparison} shows the training time comparison between the two models
% and Section \ref{sec:largerdataset} shows the result of training on a larger dataset. 

\subsection{Comparison between Models}
\label{sec:models_comparison}
We first compare various aspects of the models described in section \ref{sec:design}, including their accuracy for regression tasks, their generalizability to CPU time prediction and to classification tasks, and their training and inference speed. %This subsection tries to investigate 
These correspond to research question 1 and research question 2 from Section ~\ref{sec:research_question}.

\begin{figure*}
    % \centering
    \begin{subfigure}[b]{0.33\textwidth}
        \includegraphics[width=\textwidth]{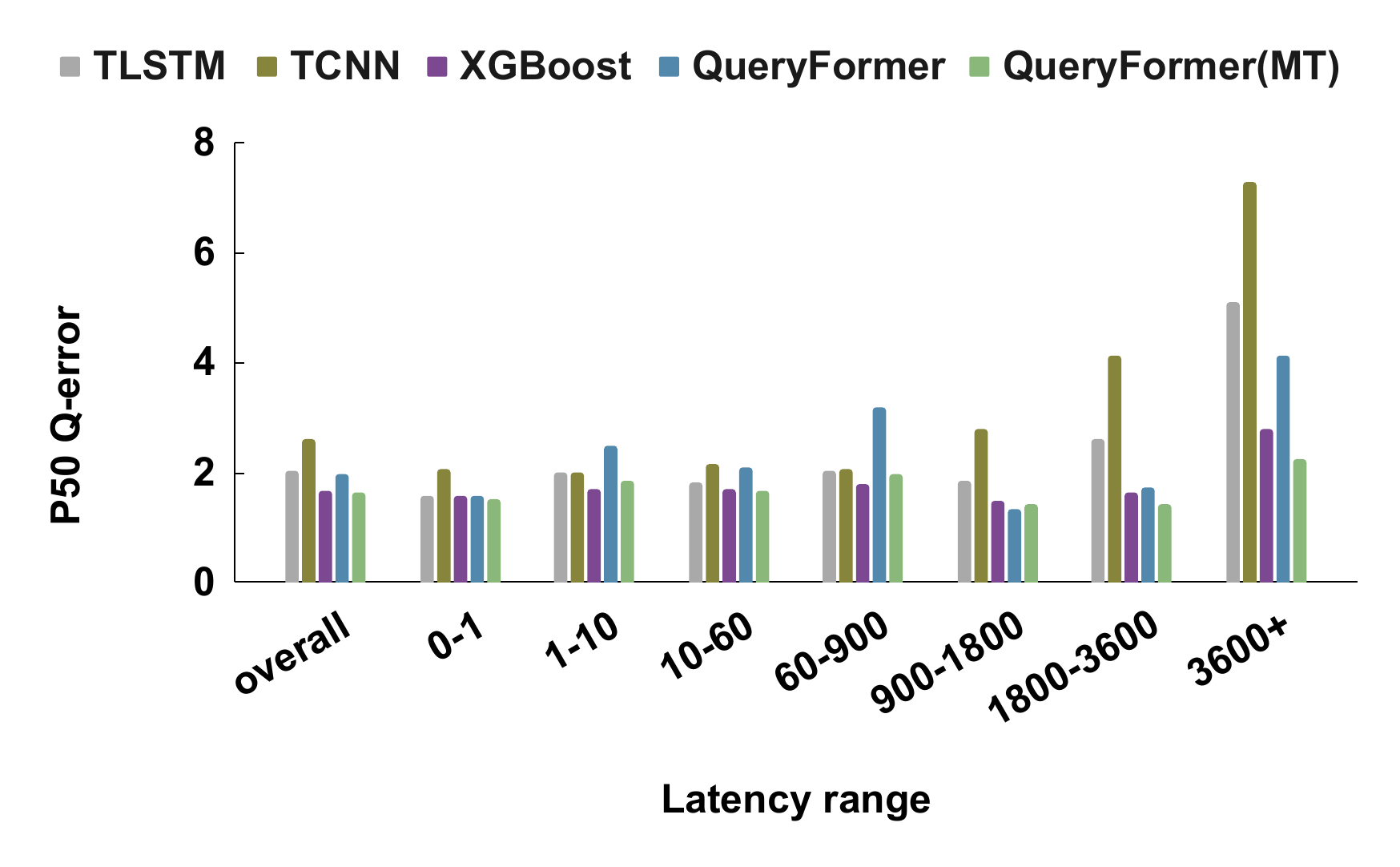}
        \caption{Comparison on P50 quantile Q-error on latency regression task for Trino}
        \label{fig:p50_qerror_latency}
    \end{subfigure}
    \hfill
    \begin{subfigure}[b]{0.33\textwidth}
        \includegraphics[width=\textwidth]{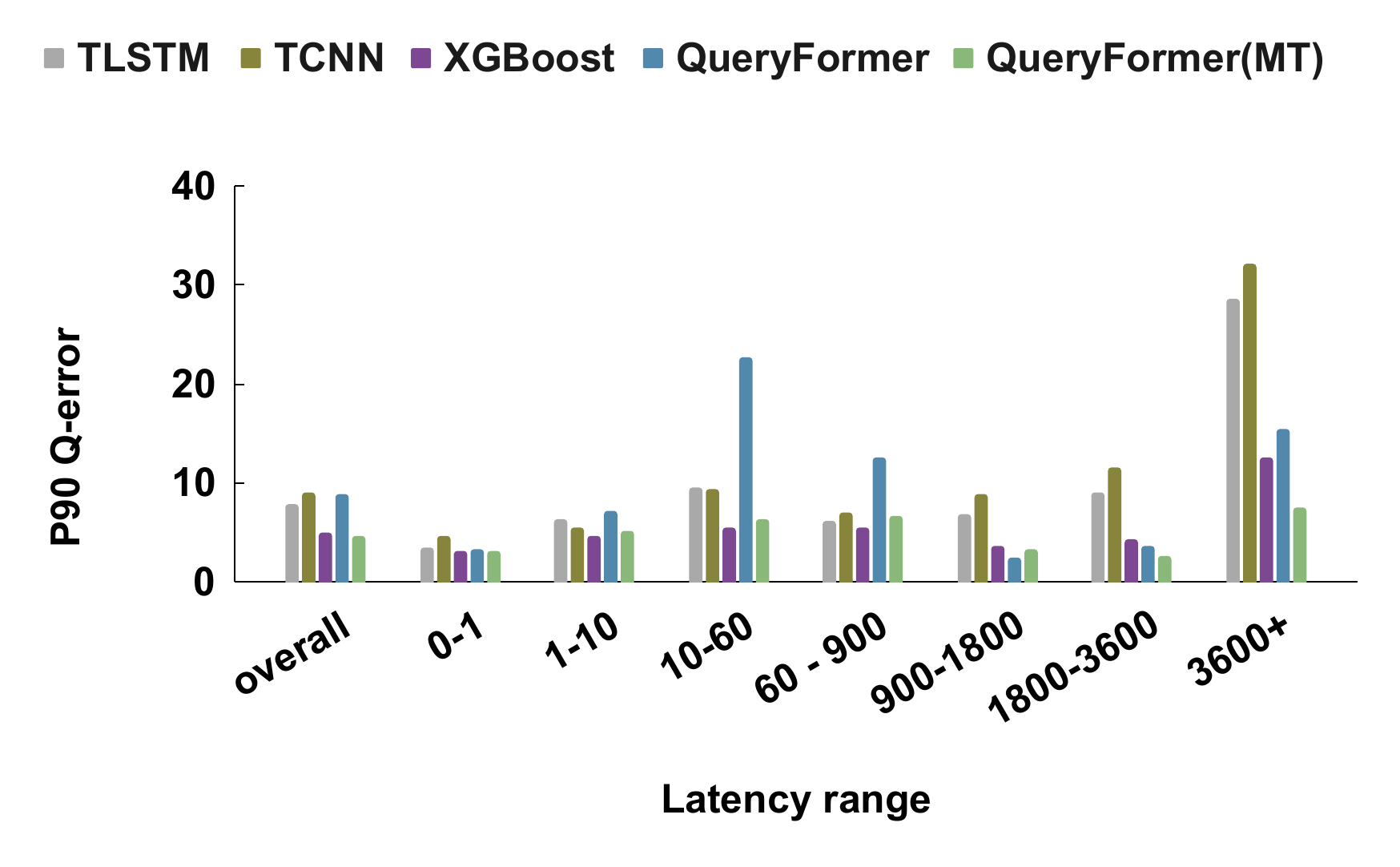}
        \caption{Comparison on P90 quantile Q-error on latency regression task for Trino}
        \label{fig:p90_qerror_latency}
    \end{subfigure}
    \hfill
    \begin{subfigure}[b]{0.33\textwidth}
        \includegraphics[width=\textwidth]{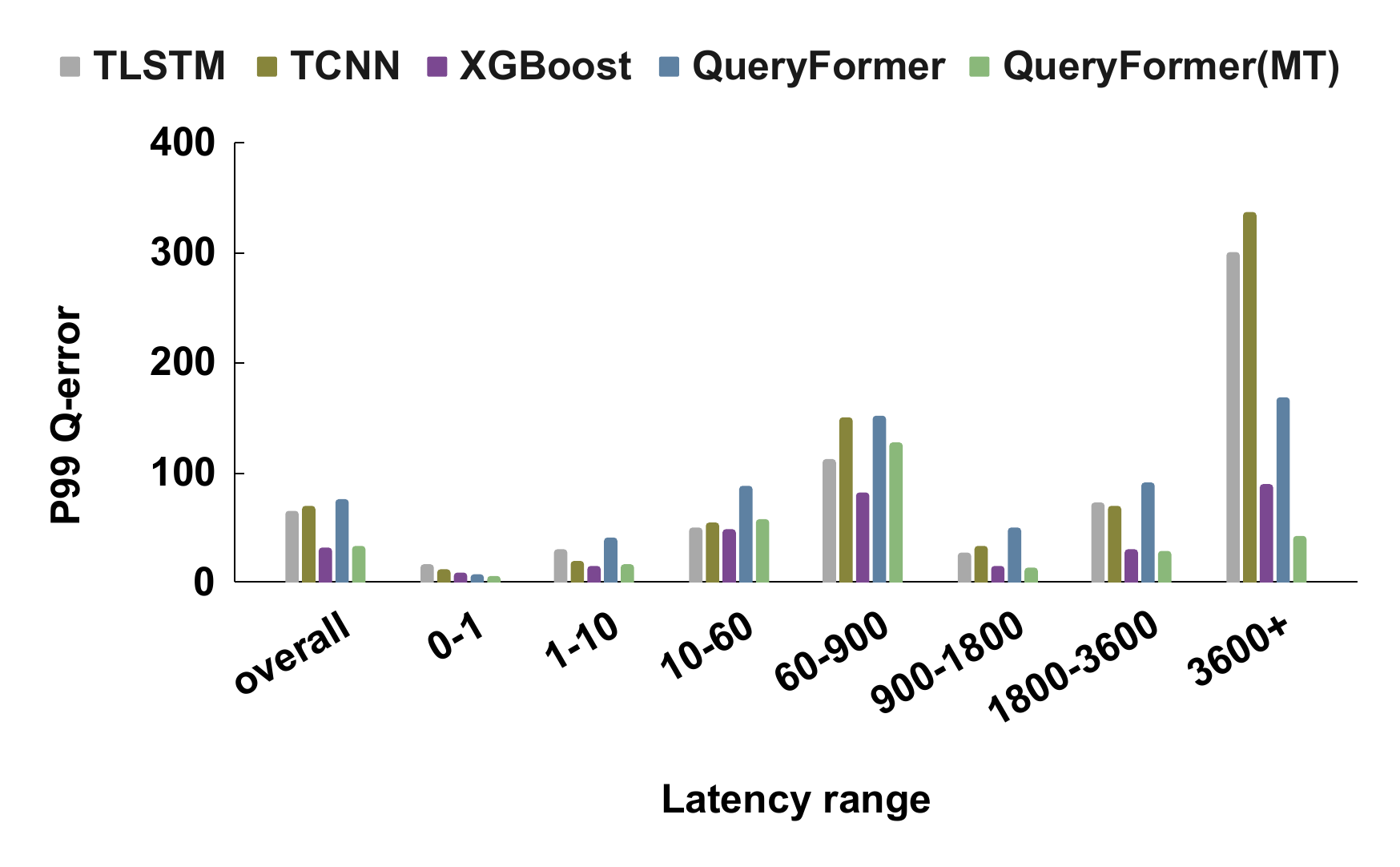}
        \caption{Comparison on quantile P99 Q-error on latency regression task for Trino}
        \label{fig:p99_qerror_latency}
    \end{subfigure}
    \caption{Comparison between model performance on latency regression task for Trino}
    \label{fig:qerror_latency}
\end{figure*}

\begin{figure*}
    \begin{subfigure}[b]{0.33\textwidth}
        \includegraphics[width=\textwidth]{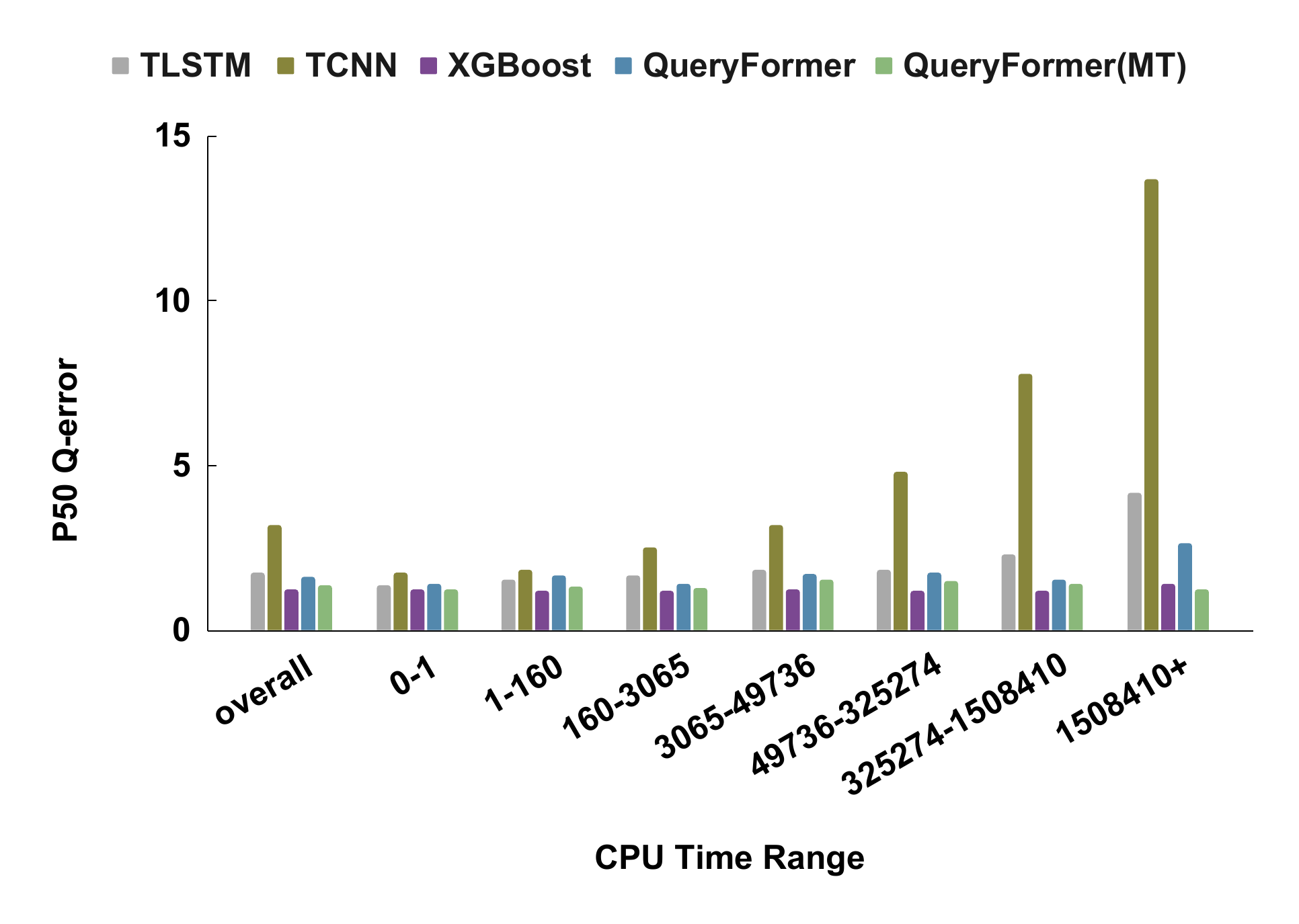}
        \caption{Comparison on P50 Q-error on CPU time regression task on Trino}
        \label{fig:p50_qerror_cpu}
    \end{subfigure}
    \hfill
    \begin{subfigure}[b]{0.33\textwidth}
        \includegraphics[width=\textwidth]{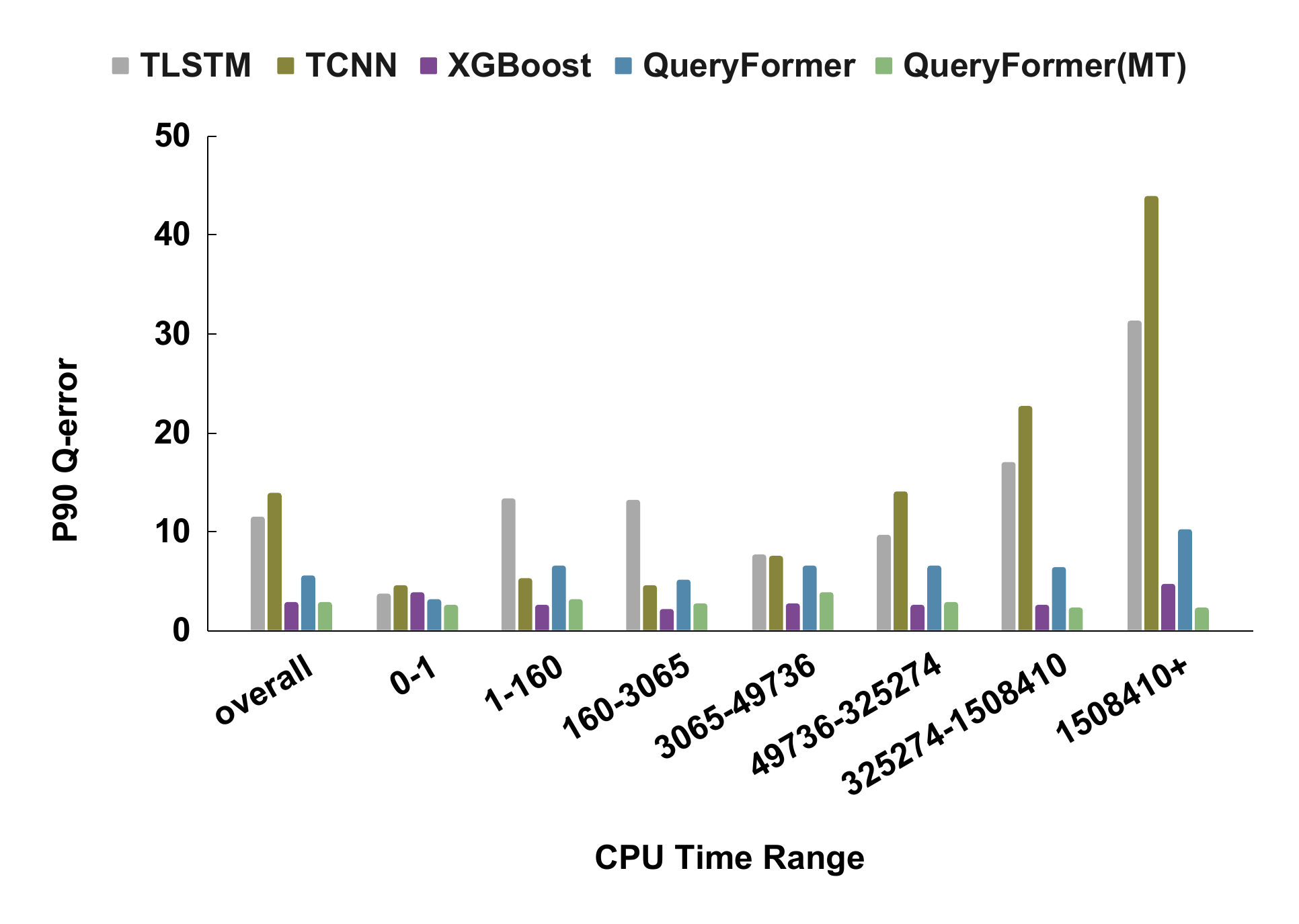}
        \caption{Comparison on P90 Q-error on CPU time regression task on Trino}
        \label{fig:p90_qerror_cpu}
    \end{subfigure}
    \hfill
    \begin{subfigure}[b]{0.33\textwidth}
        \includegraphics[width=\textwidth]{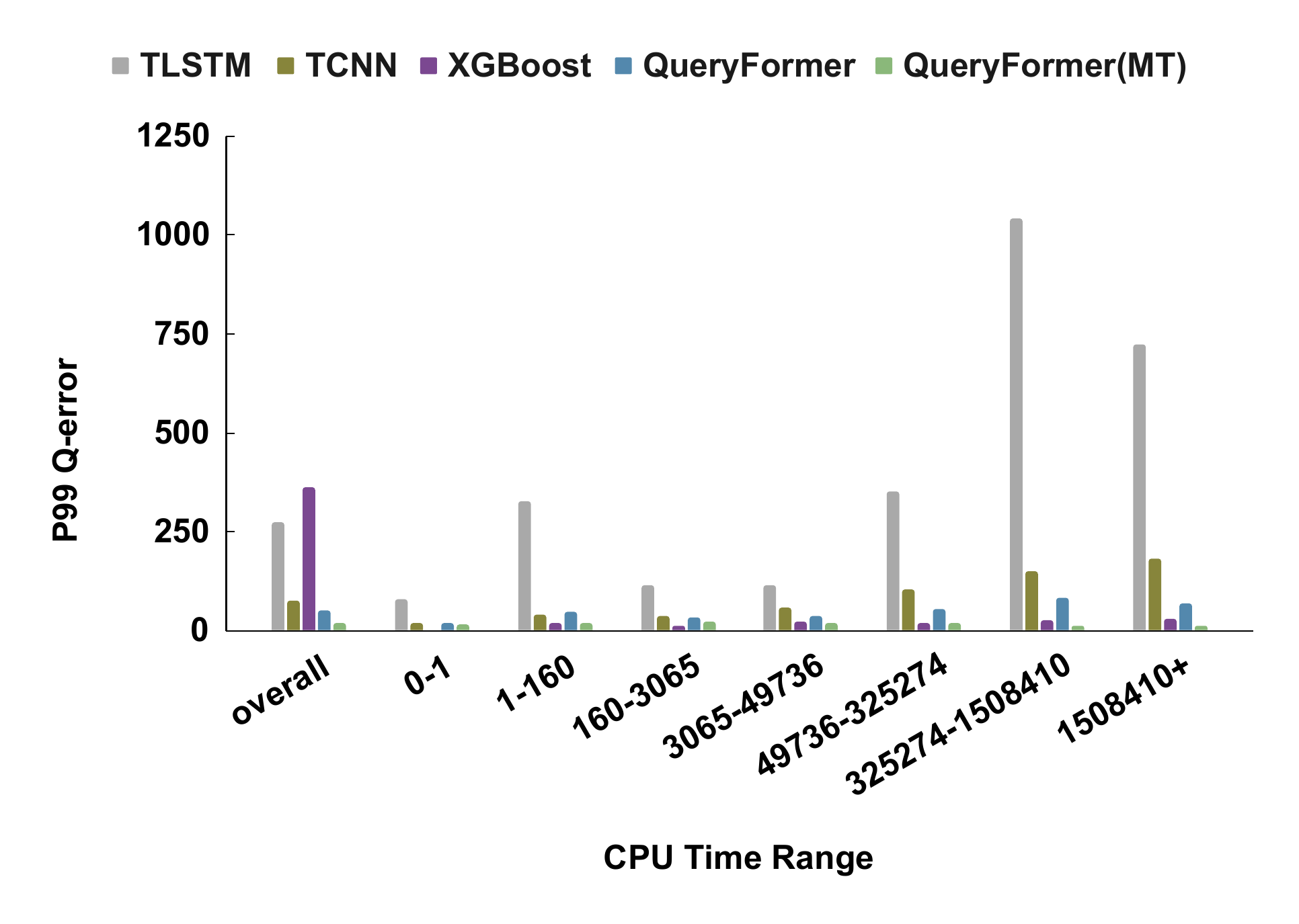}
        \caption{Comparison on P99 Q-error on CPU time regression on Trino}
        \label{fig:p99_qerror_cpu}
    \end{subfigure}
    \caption{Comparison between models on CPU Time regression task for Trino }
    \label{fig:qerror_cpu}
\end{figure*}

\subsubsection{Regression Task}
In this section, we run experiments on 4 existing models: TLSTM, TCNN, XGBoost, and Query Former. In Section \ref{sec:multitask}, we also proposed incorporating multi-task learning to improve accuracy. We implemented this within QueryFormer, and use this implementation as a fifth model in our experiments.

\textbf{Latency:}
Figure \ref{fig:qerror_latency} shows the accuracy of the compared models on a regression task that predicts latency for Trino queries. %Additionally, to explore the efficacy of our proposed multi-task learning, we incorporate QueryFormer with multi-task learning optimization enabled as the fifth model, which we 
The implementation of QueryFormer with our proposed multi-task learning addition is referred to as QueryFormer(MT) in the figure. Given the substantial variance in query latency, employing a single metric across the entire dataset fails to adequately capture performance across queries within different latency magnitudes. 
%For instance, an impressive overall performance might mask less than optimal model performance on queries with shorter or larger latencies. 
Therefore, we assess the prediction accuracy of the models across different latency ranges, where latency refers to the \textbf{actual} execution time of the queries.

The x-axis in the figure denotes latency ranges, with intervals such as 0 - 1  representing queries within that specific latency bracket. The q-error, 
%adhering to the definition provided 
as defined in the literature ~\cite{DBLP:journals/pvldb/SunL19, DBLP:journals/pvldb/ZhaoCSM22, DBLP:journals/pvldb/MarcusNMZAKPT19}, is calculated as follows:

\begin{equation}
Q\text{-}error =\frac{\max(y_{pred}, y_{label})}{\min(y_{pred}, y_{label})}
\end{equation}

Q-error has shown to be directly connected to plan optimality and plan cost
~\cite{DBLP:journals/pvldb/MoerkotteNS09}.

Figure \ref{fig:p50_qerror_latency}, \ref{fig:p90_qerror_latency}, and \ref{fig:p99_qerror_latency} present the $P_{50}$, $P_{90}$, and $P_{99}$ quantiles of q-error comparisons at various latency ranges. These figures show that, in terms of overall performance, QueryFormer(MT), which incorporates multi-task learning optimization, achieves the highest performance metrics, followed by XGBoost. 
%This also indicates that QueryFormer(MT) outperforms the original QueryFormer, in which multi-task learning optimization is not activated. Across the board, the 
These two models exhibit comparable performance on smaller queries. However, for larger queries, QueryFormer(MT) 
achieved better accuracy.
%demonstrates superior effectiveness. 
Moreover, QueryFormer(MT) maintained consistent performance across different latency ranges.
%, underscoring its robustness and adaptability to varying query sizes. 

\noindent\textbf{CPU Time:}
Figure \ref{fig:qerror_cpu} presents a comparative analysis of the same models for a CPU time regression task for Trino, spanning various CPU time ranges. The classification of CPU time intervals is determined by the corresponding quantiles of the class for latency, ensuring that each latency or CPU time range encompasses an equivalent number of samples.

%The analysis derived from these figures indicates that 
Similar to the previous set of results, QueryFormer(MT) and XGBoost again achieve the highest prediction accuracy, with QueryFormer(MT) the slightly more accurate of the two of them.  
%achieves the highest performance metrics in predicting CPU time, with XGBoost trailing closely behind. Notably, 
The prediction accuracy for CPU time is overall higher relative to the prediction for latency.
%, a trend particularly pronounced in the $P_{90}$ quantile q-error, which not only verifies the models' generalizability to CPU time prediction but also shows that CPU time is a more robust metric compared to latency , as it is largely 
This is because CPU time is more directly dependent on the query itself, and is more resilient to variations in the execution environment, especially in multi-tenant environments where the workload of other systems can fluctuate. Latency 
%is more sensitive to I/O time changes compared to CPU time. 
can be impacted by transient I/O bottlenecks. For example, queries that need to scan a table stored in Hive with thousands of partitions rely on the Hive Metastore to return relevant metadata about those partitions.
%the information of such partitions. 
For tables with many partitions, this 
process can become a bottleneck. When other queries concurrently interact with the same Hive Metastore, this bottleneck is magnified 
%operation can easily become a bottleneck, slowing down the overall query execution time. Furthermore, considering that other queries will also interact with Hive Metastore, the execution time of these queries will also be affected
~\cite{DBLP:conf/cloud/LiaoA23}. In contrast, CPU time
%which represents the amount of computation needed to execute a query, 
is not affected by such variance in I/O performance.

Both XGBoost and QueryFormer(MT) 
exhibit consistent performance across different CPU time ranges, indicating that these models generalize well to predicting CPU time in addition to latency, even across large ranges.
%predictionsuggesting their robust capability in predicting CPU time across large ranges.

\subsubsection{Classification Task}

\begin{table*}[h]
\centering
\begin{adjustbox}{max width=\textwidth}
\begin{tabular}{cccccccccccccccc}
\toprule
& \multicolumn{5}{c}{Precision} & \multicolumn{5}{c}{Recall} & \multicolumn{5}{c}{F1} \\
\cmidrule(lr){2-6} \cmidrule(lr){7-11} \cmidrule(lr){12-16}
Class & TLSTM & TCNN & XGB & QF & QF(MT)
& TLSTM & TCNN & XGB & QF & QF(MT)
& TLSTM & TCNN & XGB & QF & QF(MT) \\
\midrule
\multicolumn{16}{c}{Latency Classification} \\
\midrule
1& 0.54 & 0.68 & \textbf{0.69} & 0.53 & 0.58 & \textbf{0.99} & 0.93 & 0.98 & \textbf{0.99} & \textbf{0.99} & 0.70 & 0.79 & \textbf{0.81} & 0.69 & 0.73 \\
2&0.42 &	0.69&	\textbf{0.74}&	0.43&	0.59& 0.17&	0.48&	\textbf{0.50}&	0.11&	0.31& 0.24&	0.57&	\textbf{0.60}&	0.18&	0.41\\
3 &0.66&	0.70&	\textbf{0.73}&	0.51&	0.72& 0.58&	0.72&	\textbf{0.75}&	0.72&	0.61& 0.62&	0.71&	\textbf{0.74}&	0.60&	0.66\\
4&0.32&	\textbf{0.56}&	0.54&	0.20&	0.46& 0.58&	0.65&	\textbf{0.72}&	0.53&	0.63& 0.41&	0.60&	\textbf{0.62}&	0.29&	0.53\\
5&0.33&	0.41&	\textbf{0.43}&	0.00&	0.34& 0.40&	\textbf{0.62}&	0.47&	0.00&	0.61& 0.36&	\textbf{0.49}&	0.45&	-&	0.44\\
6&0.29&	0.42&	\textbf{0.52}&	0.00&	0.20& 0.06&	0.25&	\textbf{0.39}&	0.00&	0.03& 0.10&	0.31&	\textbf{0.45}&	-&	0.05\\
7&0.00&	\textbf{0.93}&	0.87&	0.00&	0.89& 0.00	&0.17&	\textbf{0.29}&	0.00&	0.00& -&	0.29&	\textbf{0.44}&	-&	0.00\\
\midrule
\multicolumn{16}{c}{CPU Time Classification} \\
\midrule
1& 0.85&	0.91&	\textbf{0.93}&	0.79&	0.85& \textbf{0.98}&	0.94&	\textbf{0.98}&	\textbf{0.98}&	\textbf{0.98}& 0.91 &	0.92&	\textbf{0.95}&	0.87&	0.91\\
2&0.68&	0.89&	\textbf{0.94}&	0.48&	0.83& 0.79&	0.84&	\textbf{0.91}&	0.73&	0.80& 0.73&	0.86&	\textbf{0.92}&	0.58&	0.81\\
3&0.77&	0.86&	\textbf{0.89}&	0.44&	0.73& 0.66&	0.87&	\textbf{0.92}&	0.39&	0.85& 0.71&	0.86&	\textbf{0.90}&	0.41&	0.79\\

4& 0.48&	0.78&	\textbf{0.84}&	0.20&	0.67& 0.63&	0.79&	\textbf{0.84}&	0.51&	0.64& 0.54&	0.78&	\textbf{0.84}&	0.29&	0.65\\

5&0.58&	0.77&	\textbf{0.86}&	0.00&	0.33& 0.39&	0.75&	\textbf{0.85}&	0.00&	0.72& 0.47	&0.76&	\textbf{0.85}&	-&	0.45\\

6&0.46&	0.58&	\textbf{0.77}&	0.00&	0.00& 0.62&	\textbf{0.86}&	0.83&	0.00&	0.00& 0.53&	0.69&	\textbf{0.80}&	-&	-\\

7 & 0.00&	\textbf{0.85}&	0.84&	0.00&	0.00& 0.00&	0.21&	\textbf{0.66}&	0.00&	0.00 &-	&0.34&	\textbf{0.74}&	-	&-\\

\bottomrule
\end{tabular}
\end{adjustbox}
\caption{Comparison between models on classification task on latency and CPU time}
\label{table:classification_comparison}
\end{table*}

Section \ref{sec:classification_task}, described some
%there are instances 
cases where the objective is to classify queries into specific categories based on their characteristics. This section investigates the generalizability of the classification task. 

Table \ref{table:classification_comparison} presents the precision, recall, and F1 score of the models when placing queries into one of 7 latency classes and 7 CPU Time classes. 
%prediction across different models on 
%different classes.
The seven latency classes used for this table correspond to the seven ranges from Figure \ref{fig:qerror_latency}, and the seven CPU time classes correspond to the seven ranges from Figure \ref{fig:qerror_cpu}.
%The assignment of classes is based on the same criteria for assigning ranges for the regression task. Overall, XGBoost emerges as the top performer among the models evaluated, realizing the 
XBoost yields the highest performance in all but class 5 for latency classification. %A notable disparity observed is that certain models, such as 
TLSTM and QueryFormer both fail to accurately predict queries within the larger classes, as evidenced by their 0\% precision and recall rates in predicting large queries. 
%Regarding recall, a similar trend is observed; however, it's worth mentioning that 
All models %demonstrate exceptionally 
achieve high recall rates for small queries. 
%Overall, the XGBoost model generalizes better in terms of the classification task compared to neural-network-based approaches. 

These results also indicate that classifying queries by CPU time is more accurate than classifying them by latency. This is likely due to the sames reasons discussed in the previous section, that there are more external factors that can influence latency, and therefore any kind of prediction related to latency is harder than prediction related to CPU time.

\subsubsection{Training and Inference Time Comparison}
\label{sec:timecompare}

Table \ref{tab:comparison_training_inference_time} provides a comparison of the training and inference times for these models tested on a dataset comprising 35,050 samples. The training time is measured from the start of training until the time when the stop condition is met, as defined in Section \ref{sec:expsetup}
. 
%In terms of inference speed, the time taken to perform inference on the test dataset is evaluated.

%The comparison reveals that 
XGBoost's training time was 1-2 orders of magnitude faster than the neural network models, and 2-3 orders of magnitude faster for inference on the test dataset. 
%boasting a 220x acceleration in training and a remarkable 1852x speed-up in inference relative to QueryFormer(MT). 
For the QueryFormer(MT) model, inference is conducted in batches of 64, resulting in a per-batch inference time of 0.77 seconds and a per-sample inference time of 0.01 seconds. Its batch-based processing model facilitates parallel processing, which improves scalability, but it is
%However, it's important to note that the computation speed-up for neural networks benefits substantially from data parallelism inherent in batch processing, which implies a potentially higher real per-sample inference time. This aspect could render neural network-based models 
less suitable for an individual low-latency interactive query. 
% since the parallel processing can introduce additional latency to the overall end-to-end response time. 
%On the contrary, XGBoost does not share this limitation, presenting a more favorable option for scenarios demanding 
XGBoost thus has an additional advantage for workloads that involve
highly time-sensitive query processing.

Neural network models also require hyperparameter tuning, which is an inherently sequential process, that can further increase
%that can easily multiply the end-to-end time until the model is ready for inference, exaggerating the problem of prolonged 
training time.

\begin{table}[h]
\centering
\caption{Comparison between models on training and inference time}
\begin{tabularx}{\linewidth}{@{} 
>{\hsize=.27\hsize\centering\arraybackslash}X
>{\hsize=.14\hsize\centering\arraybackslash}X
>{\hsize=.14\hsize\centering\arraybackslash}X
>{\hsize=.12\hsize\centering\arraybackslash}X
>{\hsize=.14\hsize\centering\arraybackslash}X
>{\hsize=.19\hsize\centering\arraybackslash}X
@{}}
\toprule
& \textbf{TLSTM} & \textbf{TCNN} & \textbf{XGBoost} & \textbf{QF} & \textbf{QF(MT)} \\
\midrule
\textbf{Training [s]} & 27000 & 3000 & \textbf{278} & 32400 & 61200 \\
\textbf{Inference [s]} & 270 & 32 & \textbf{0.2} & 477 & 426 \\
\bottomrule
\end{tabularx}
\label{tab:comparison_training_inference_time}
\end{table}

\subsection{Multi-task Learning}
\label{sec:multi_task_res}
\begin{figure}[h]
        \centering
        \includegraphics[width=\linewidth]{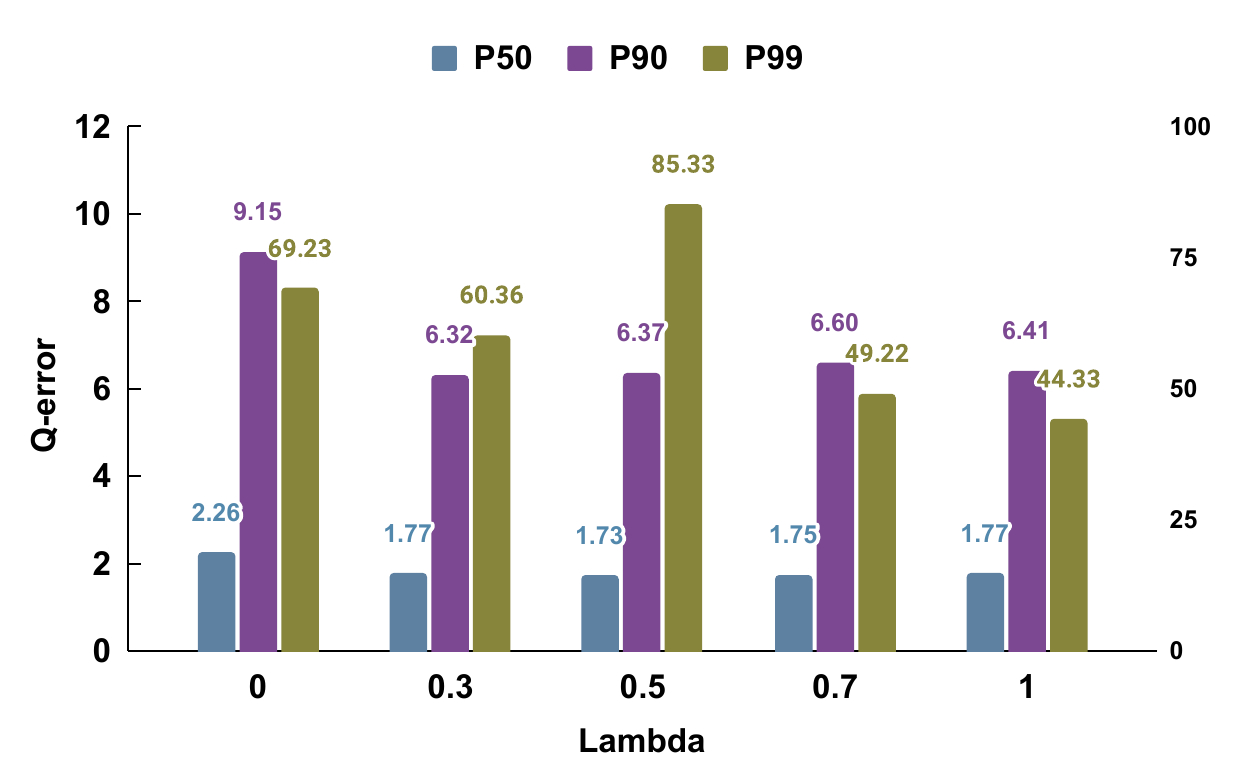}
        \caption{Effect of parameter $\lambda$ on multi-task learning performance}
        \label{fig:mt_parameter_tuning}
\end{figure}

%This subsection tries to explore the 
We now investigate the
effectiveness of the multi-task learning optimization
(research question 3 from Section ~\ref{sec:research_question}).
Figure \ref{fig:mt_parameter_tuning} examines the influence of the $\lambda$ parameter when using multi-task learning in 
 %multi-task learning optimization for 
 QueryFormer. The $P_{50}$ and $P_{90}$ quantile q-errors use the numbers from the left axis, while the $P_{99}$ quantile Q-error is plotted against the right axis.

The figure shows that multi-task learning (scenarios where $\lambda$ is greater than 0), significantly enhances performance over the baseline scenario where multi-task learning is not employed ($\lambda = 0$). 
The outcomes for both $P_{50}$ and $P_{90}$ quantile q-errors remain steady across varying $\lambda$ values; however, the performance peaks for the P99 quantile q-error when $\lambda$ is set to 1.
%, surpassing performance metrics under other settings.

\subsection{Classification Task Number of Classes}
\label{sec:num_class}

Table \ref{table:classification_num_classes} examines how latency and CPU time classification accuracy is affected when reducing the number of classification classes from 7 to  4, and 2 
%respectively—on the accuracy of latency and CPU time classification 
for Trino queries using XGBoost. 
%This reduction showcases a clear trend: a
As the number of classes decreases, the classification accuracy correspondingly increases. Given that binary classification suffices for a variety of applications, this improved accuracy can yield significant impact. 
%this pattern could significantly facilitate such tasks. 
The observed trend towards higher accuracy with fewer classification categories can be attributed to the increased consistency in query labeling, which we analyze in detail in Section \ref{sec:analysis}. 

%Consistent with previous results, classification on CPU time achieves better performance.

For classification tasks involving a large number of classes, the distinction between a correct prediction, a prediction that is off by one class, and a prediction that is significantly incorrect becomes important. In scenarios where strict accuracy is not mandatory, a prediction that is one class off may be acceptable. To measure the models' performance under this rationale, we modified the calculation of precision and recall to include predictions that are off by one class as correct, and measure the models' performance under this rationale. Specifically, if a data point's predicted class is \( x \) and its true class is \( x+1 \) or \( x-1 \), we count it as a correct prediction for class \( x \). The results shown in the columns labeled \textbf{7 classes mix} reflect this adjusted calculation. 
%Comparing these results with those under the columns labeled \textbf{7 classes} we observe that even for the 7-class scenario, the models tend to make errors that are only one class off from the true label. This is particularly evident in the 0.99 accuracy result for CPU time prediction.
All metrics improve significantly relative 
to the original \textbf{7 classes} numbers, indicating that most cases of incorrect classification are indeed only one class away from the correct classification. The 0.99 accuracy result for CPU time prediction is particularly notable.

% \begin{table*}[h]
% \centering
% \begin{adjustbox}{max width=\textwidth}
% \begin{tabular}{cccccccccccccc}
% \toprule
% & \multicolumn{7}{c}{\textbf{7 classes}} &
% \multicolumn{4}{c}{\textbf{4 classes}} & \multicolumn{2}{c}{\textbf{2 classes}} \\
% \cmidrule(lr){2-8} \cmidrule(lr){9-12} \cmidrule(lr){13-14}
%  & 1 & 2 & 3 & 4 & 5
% & 6 & 7 & 1 & 2 & 3
% & 4 & 1 & 2   \\
% \midrule
% \multicolumn{14}{c}{\textbf{Latency Classification}} \\
% \midrule
% Precision& 0.69&0.74&	0.73&	0.54&	0.43&	0.52&	0.87&	0.86&	0.84&	0.64&	0.85&	0.94&	0.96\\
% Recall&0.98&	0.5&	0.75&	0.72&	0.47&	0.39&	0.29&	0.97&	0.78&	0.78&	0.62&	0.96&	0.94\\
% F1 Score & 0.81	 & 0.6	 & 0.74	 & 0.62	 & 0.45	 & 0.45	 & 0.43	 & 0.91	 & 0.81 & 	0.7	 & 0.71	 & 0.95	 & 0.95\\
% Accuracy &\multicolumn{7}{c}{0.62} & \multicolumn{4}{c}{0.79} &
% \multicolumn{2}{c}{0.95}\\
% \midrule
% \multicolumn{14}{c}{\textbf{CPU Time Classification}} \\
% \midrule
% Precision& 0.93& 	0.94& 	0.89& 	0.84& 	0.86& 	0.77& 	0.84& 0.96 &	0.93& 	0.87& 	0.95& 	0.97& 	0.98 \\
% Recall& 0.98& 	0.91& 	0.92& 	0.84& 	0.85& 	0.83& 	0.66& 0.98 &	0.91& 	0.91& 	0.9& 	0.98& 	0.97  \\
% F1 Score & 0.96& 	0.92& 	0.91& 	0.84& 	0.86& 	0.8& 	0.74&0.97 & 	0.92& 	0.89& 	0.92& 	0.97& 	0.97 \\
% Accuracy &\multicolumn{7}{c}{0.87} & \multicolumn{4}{c}{0.92} &
% \multicolumn{2}{c}{0.97}\\
% \bottomrule
% \end{tabular}
% \end{adjustbox}
% \caption{Effect on the number of classes on classification task result}
% \label{table:classification_num_classes}
% \end{table*}

\begin{table*}[h]
\centering
\begin{adjustbox}{max width=\textwidth}
\begin{tabular}{cccccccccccccccccccccc}
\toprule
& \multicolumn{7}{c}{\textbf{7 classes}} &
\multicolumn{7}{c}{\textbf{7 classes mix}} &
\multicolumn{4}{c}{\textbf{4 classes}} & \multicolumn{2}{c}{\textbf{2 classes}} \\
\cmidrule(lr){2-8} \cmidrule(lr){9-15} \cmidrule(lr){16-19} \cmidrule(lr){20-21}
 & 1 & 2 & 3 & 4 & 5
& 6 & 7 &1 & 2 & 3 & 4 & 5
& 6 & 7& 1 & 2 & 3
& 4 & 1 & 2   \\
\midrule
\multicolumn{21}{c}{\textbf{Latency Classification}} \\
\midrule
Precision& 0.69&0.74&	0.73&	0.54&	0.43&	0.52&	0.87&0.59&0.82&0.79&0.55&0.45&0.87&0.97&	0.86&	0.84&	0.64&	0.85&	0.94&	0.96\\
Recall&0.98&	0.5&	0.75&	0.72&	0.47&	0.39&	0.29&1.00&0.99&0.89&0.90&0.97&0.80&0.19&	0.97&	0.78&	0.78&	0.62&	0.96&	0.94\\
F1 Score & 0.81	 & 0.6	 & 0.74	 & 0.62	 & 0.45	 & 0.45	 & 0.43&0.74&0.90&0.84&0.69&0.62&0.84&0.31 & 0.91	 & 0.81 & 	0.7	 & 0.71	 & 0.95	 & 0.95\\
Accuracy &\multicolumn{7}{c}{0.62} & 
\multicolumn{7}{c}{0.87} & 
\multicolumn{4}{c}{0.79} &
\multicolumn{2}{c}{0.95}\\
\midrule
\multicolumn{21}{c}{\textbf{CPU Time Classification}} \\
\midrule
Precision& 0.93& 	0.94& 	0.89& 	0.84& 	0.86& 	0.77& 	0.84&0.94&0.95&0.90&0.86&0.88&0.80&0.88& 0.96 &	0.93& 	0.87& 	0.95& 	0.97& 	0.98 \\
Recall& 0.98& 	0.91& 	0.92& 	0.84& 	0.85& 	0.83& 	0.66&1.00& 1.00& 1.00& 0.98& 0.98& 0.97& 0.95& 0.98 &	0.91& 	0.91& 	0.9& 	0.98& 	0.97  \\
F1 Score & 0.96& 	0.92& 	0.91& 	0.84& 	0.86& 	0.8& 	0.74&0.97&0.97&0.95&0.91&0.93&0.88&0.92&0.97 & 	0.92& 	0.89& 	0.92& 	0.97& 	0.97 \\
Accuracy &\multicolumn{7}{c}{0.87} & 
\multicolumn{7}{c}{0.99} & 
\multicolumn{4}{c}{0.92} &
\multicolumn{2}{c}{0.97}\\
\bottomrule
\end{tabular}
\end{adjustbox}
\caption{Effect on the number of classes on classification task result}
\label{table:classification_num_classes}
\end{table*}

% \begin{table}[h]
% \centering
% \caption{Trino queries latency classification task on 7 classes using XGBoost}
% \begin{tabularx}{\linewidth}{@{} *{4}{>{\centering\arraybackslash}X} @{}}
% \toprule
% \makecell{class} & \makecell{Precision} & \makecell{Recall} & \makecell{F1 Score} \\
% \midrule
% 0 & 0.93 & 0.98 & 0.96 \\
% 1 & 0.94 & 0.91 & 0.92 \\
% 2 & 0.89 & 0.92 & 0.91 \\
% 3 & 0.84 & 0.84 & 0.84 \\
% 4 & 0.86 & 0.85 & 0.86 \\
% 5 & 0.77 & 0.83 & 0.80 \\
% 6 & 0.84 & 0.66 & 0.74 \\
% \bottomrule
% \end{tabularx}
% \label{tab:7_classes}
% \end{table}

% \begin{table}[h]
% \centering
% \caption{Trino queries latency classification task on 4 classes using XGBoost}
% \begin{tabularx}{\linewidth}{@{} *{4}{>{\centering\arraybackslash}X} @{}}
% \toprule
% \makecell{class} & \makecell{Precision} & \makecell{Recall} & \makecell{F1 Score} \\
% \midrule
% 0 & 0.96 & 0.98 & 0.97 \\
% \makecell{\( 1 \)} & 0.93  & 0.91 & 0.92 \\
% \makecell{\( 2 \)} & 0.87 & 0.91 & 0.89 \\
% \makecell{\( 3 \)} & 0.95 & 0.90 & 0.92 \\
% \bottomrule
% \end{tabularx}
% \label{tab:4_classes}
% \end{table}

% \begin{table}[h]
% \centering
% \caption{Trino queries latency classification task on 2 classes using XGBoost}
% \begin{tabularx}{\linewidth}{@{} *{4}{>{\centering\arraybackslash}X} @{}}
% \toprule
% \makecell{class} & \makecell{Precision} & \makecell{Recall} & \makecell{F1 Score} \\
% \midrule
% 0 &  0.97 & 0.98 & 0.97 \\
% \makecell{\( 1 \)} & 0.98  & 0.97 & 0.97  \\
% \bottomrule
% \end{tabularx}
% \label{tab:2_classes}
% \end{table}

\subsection{Large Dataset}
\label{sec:largerdataset}
\begin{figure*}
    % \centering
    \begin{subfigure}[b]{0.24\textwidth}
        \includegraphics[width=\textwidth]{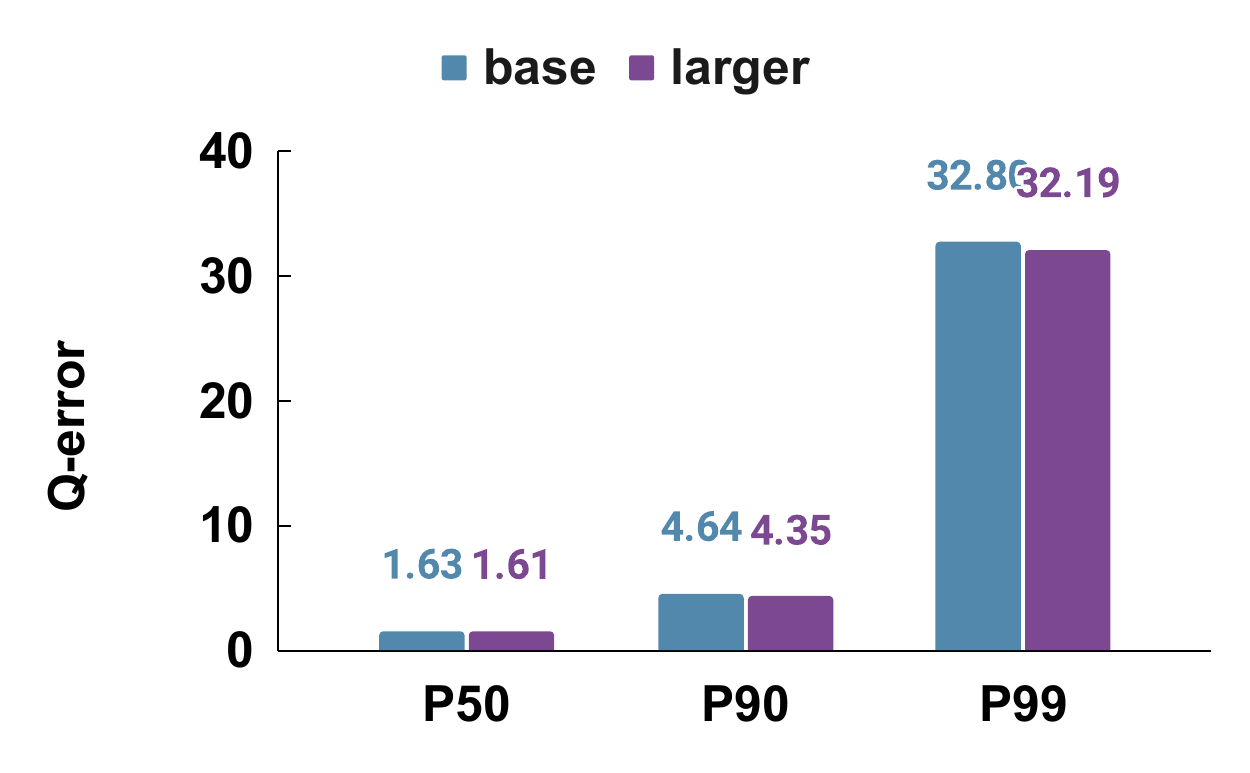}
        \caption{Result on latency prediction for QueryFormer(MT)}
        \label{fig:larger_qf_latency}
    \end{subfigure}
    \hfill
    \begin{subfigure}[b]{0.24\textwidth}
        \includegraphics[width=\textwidth]{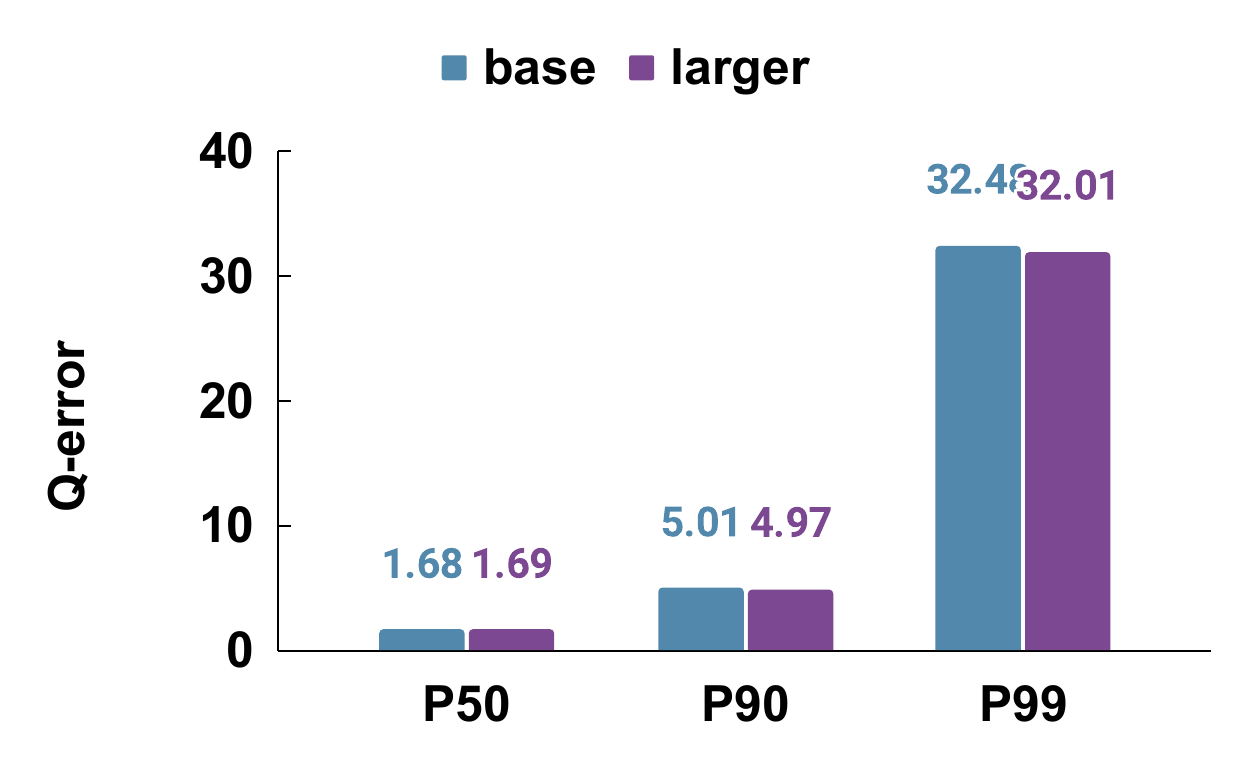}
        \caption{Result on latency prediction for XGBoost}
        \label{fig:larger_xgb_latency}
    \end{subfigure}
    \hfill
    \begin{subfigure}[b]{0.24\textwidth}
        \includegraphics[width=\textwidth]{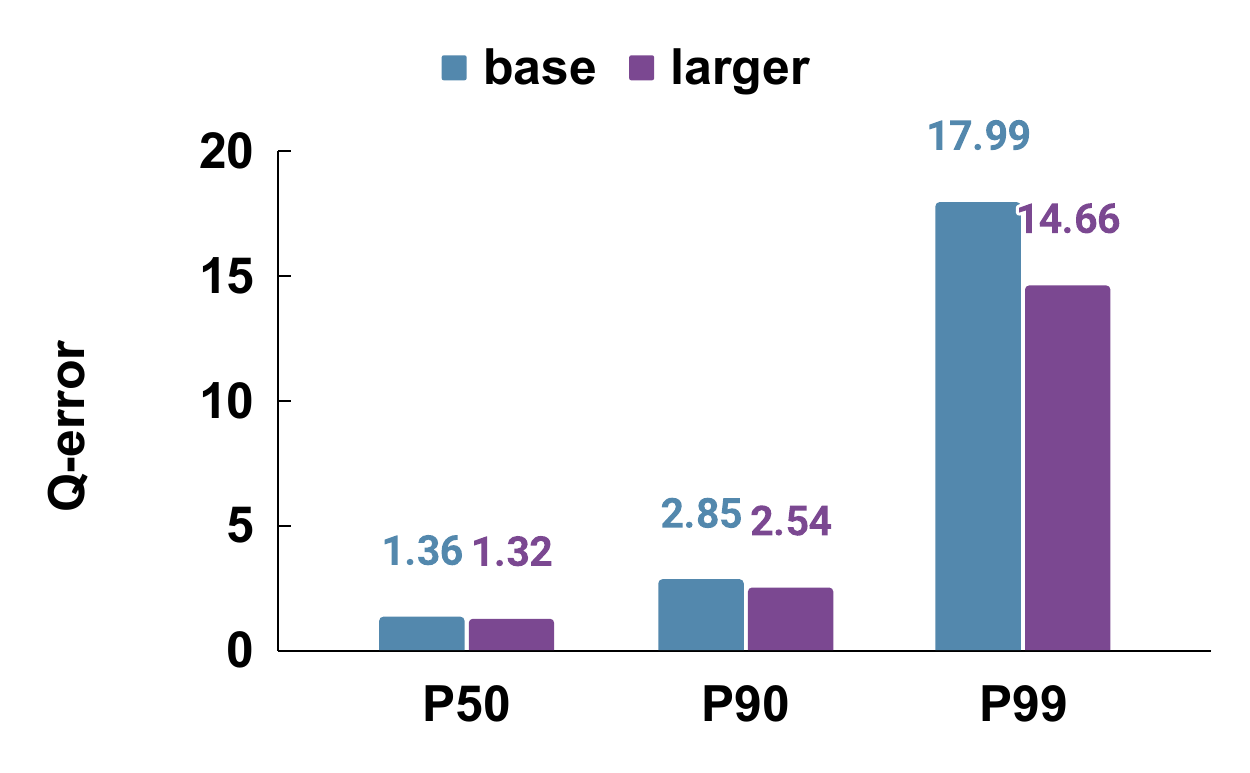}
        \caption{Result on CPU time prediction for QueryFormer(MT)}
        \label{fig:larger_qf_cpu}
    \end{subfigure}
    \hfill
    \begin{subfigure}[b]{0.24\textwidth}
        \includegraphics[width=\textwidth]{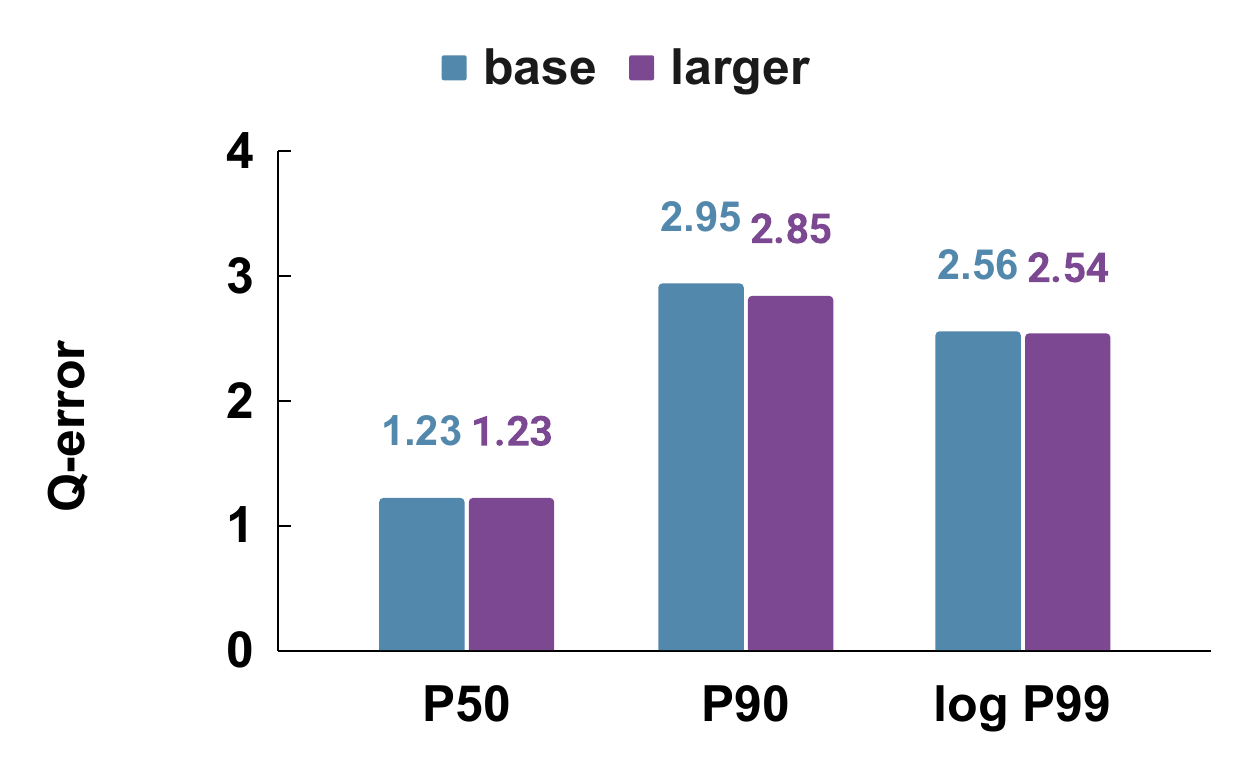}
        \caption{Result on CPU time prediction for XGBoost}
        \label{fig:larger_xgb_cpu}
    \end{subfigure}
    \caption{Performance of QueryFormer(MT) and XGBoost on a larger dataset}
    \label{fig:larger_mini}
\end{figure*}

To determine whether model performance is limited by the volume of training data,
%—given the empirical consensus that more data generally leads to better performance—we take an 
we run an experiment where the size of the training dataset is doubled. Figures \ref{fig:larger_qf_latency}, \ref{fig:larger_xgb_latency}, \ref{fig:larger_qf_cpu}, and \ref{fig:larger_xgb_cpu} show the results of this experiment for QueryFormer(MT) and XGBoost, which were the models yielding the highest performance for the previous sets of experiments. Figure \ref{fig:larger_qf_cpu} employs a logarithmic scale for the $P_{99}$ quantile q-error due to its large magnitude.
%Contrary to expectations, the results indicate that 
The results indicate that increasing the dataset size does not necessarily lead to a significant improvement in prediction accuracy.  
%due to its disproportionately large magnitude. The rationale behind this potential ceiling effect where additional data fails to contribute to performance improvements, is 
This reason for this is that these models were already achieving accuracy levels close to their theoretical lower bound, which will be explained further in Section \ref{sec:analysis}.

\subsection{Summary}
We summarize our findings and address research questions 1–3 as follows:

\begin{itemize}
    \item \textbf{Model Comparison}: For regression tasks, the QueryFormer (MT) model achieves the best overall performance, maintaining consistent results across varying latency ranges. In contrast, for classification tasks, XGBoost emerges as the top-performing model, outperforming other approaches.
    \item \textbf{Effectiveness of Multi-Task Learning}: Multi-task learning proves highly effective, leveraging operator-level statistics to enhance predictive performance across multiple tasks.
    \item \textbf{Model Application on New Targets}: When deploying models on new targets, using CPU time as the target metric results in superior predictive performance compared to latency-based metrics.
    \item \textbf{Cost-Effectiveness of Models}: XGBoost demonstrates exceptional cost-effectiveness, requiring significantly less computational time and resources while maintaining acceptable levels of performance.
    \item \textbf{Impact of Model Size}: Increasing the size of a model does not necessarily lead to improved performance, indicating that the primary limitation for further model enhancement may not lie in data size.
\end{itemize}

\section{Analysis}
\label{sec:analysis}
This section investigates the theoretical performance limits of the models experimented with in the previous section given the available data (research question 4 from section ~\ref{sec:research_question}). The encoding process outlined in Section~\ref{sec:nnencoding} 
%for the original query plan 
is inherently lossy, leading to the omission of %critical 
information that could significantly influence query performance. For example, details regarding join conditions and filter predicates
%, which can profoundly impact latency and CPU time, 
are not captured in the encoding. 
%This means that two 
Two queries with identical shapes but differing in operator-specific details
%—thereby affecting their latency and CPU time—
will be represented by the same encoded input for the models, even though they may differ significantly in latency and CPU time.

\begin{figure}[h]
    \centering
    % \begin{minipage}{0.48\linewidth}
        \centering
  \includegraphics[width=\linewidth]{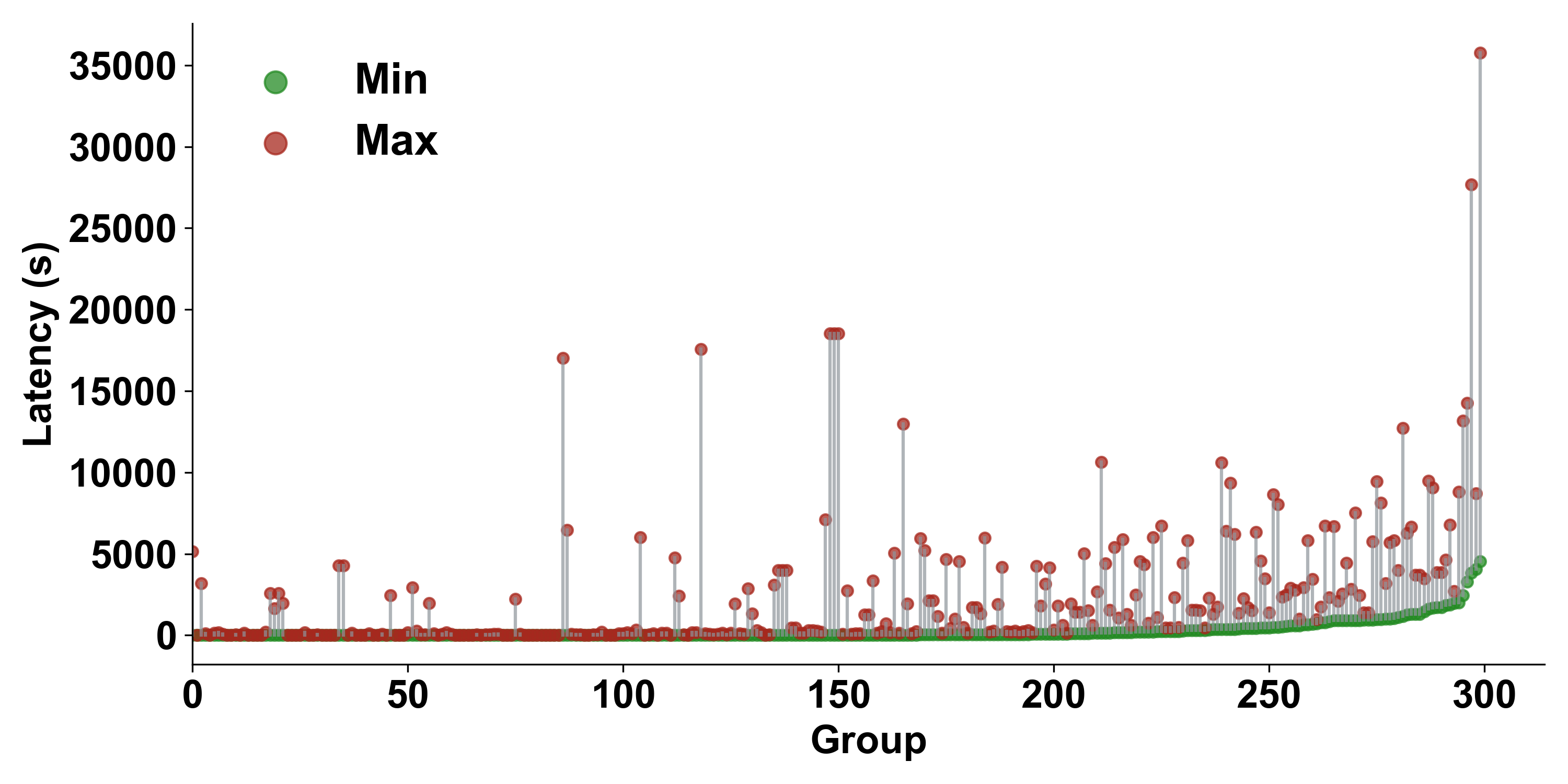}
  \caption{Latency variation within the same encoded form group}
  \label{same_group_latency}
  \Description{Number of Trino queries per day across a month}
    % \end{minipage}\hfill
  %   \begin{minipage}{0.48\linewidth}
  %       \centering
  % \includegraphics[width=\linewidth]{figures/same_group_cputime.png}
  % \caption{Cputime variation within the same group}
  % \label{same_group_latency}
  % \Description{Number of Trino queries per day across a month}
  %   \end{minipage}
\end{figure}

% We define the signature of a query as the shape of the query(as defined in section \ref{sec:datagen}) and the magnitude of input statistics. Note this signature can represent the most important information of the encoded query plan.
To illustrate this point, Figure \ref{same_group_latency} shows the variance in latency for queries with similar encoded forms. Specifically, it takes 300 encoded forms, finds all queries that are encoded with one of these 300 forms, and shows the latency differences of these queries within the same equivalent form. 
%when grouping queries according to the equality of their encoded forms for 300 distinct encoded forms.  On these graphs, the 
The x-axis represents a distinct encoded form, while the y-axis illustrates the range between the maximum and minimum true label metrics for queries that are encoded with that form.
%, highlighting the substantial latency disparities that can arise from similar structures. Note that 
CPU time variance follows a similar pattern.

Due to the lossy nature of the encoding process, the models 
%are confronted with a challenging scenario: receiving 
may receive identical inputs that correspond to different outputs. 
%This situation inherently implies that the 
Therefore, the models will never be able to predict the output for these inputs with 100\% accuracy. %In essence, this establishes 
Rather, there is an intrinsic lower bound on the q-error achievable by the models.

% \begin{figure}[h]
%   \centering
%   \includegraphics[width=\linewidth]{figures/same_group_latency.png}
%   \caption{Latency variation within the same group}
%   \label{same_group_cputime}
%   \Description{Number of Trino queries per day across a month}
% \end{figure}

% \begin{figure}[h]
%   \centering
%   \includegraphics[width=\linewidth]{figures/same_group_cputime.png}
%   \caption{Cputime variation within the same group}
%   \label{same_group_latency}
%   \Description{Number of Trino queries per day across a month}
% \end{figure}

To investigate the lower bound on prediction accuracy imposed by the constraints of lossy encoding, we formulate an optimization problem. Consider the input pairs \( \{ (x_i, y_i) \mid i = 1, \ldots, n \} \), where \(x_i\) is a query plan and \(y_i\) are the performance metrics (such as latency or CPU time) that are measured during the execution of the plan. Define \( \mathcal{F} \) as the encoding function discussed in section \ref{sec:nnencoding}.

The inputs are first partitioned based on their encoded form \( \mathcal{F}(x_i) \) into equivalence partitions \( \{ G_i = \{ (x_{ij}, y_{ij}) \mid j = 1, \ldots, n_i \} \mid i = 1, \ldots, k \} \), where \(n_i\) indicates the number of items in group \(G_i\) and $k$ is the total number of groups. Plans \(x_m\) and \(x_n\) are grouped together into \(G_i\) if \( \mathcal{F}(x_m) = \mathcal{F}(x_n) \). Within each group \(G_i\), a singular prediction \(P_i\) is generated for all plans because their encoded forms are the same. The upper bound can be calculated by finding the set of \(P_i\) that minimizes the overall $P_{50}$\footnote{Alternatively, $P_{90}$ or $P_{99}$ quantile q-error could  be used.} quantile q-error:
%, which is formalized as follows:

\begin{equation}
P_{50}\left(\left\{ \left\{ \frac{\max(y_{ij}, P_i)}{\min(y_{ij}, P_i)} \mid j = 1, \ldots, n_i \right\} \mid i = 1, \ldots, k \right\}\right)
\end{equation}

where \(P_{50}\) is the function that calculates the 50-th quantile of a list.

%Considering the complexity of directly calculating these \(P_i\) values, w
We employ a greedy strategy to calculate these \(P_i\) values. For each \(P_i\), we select it from the set of candidates \( \{ y_{ij} \mid j = 1, \ldots, n_i \} \) to minimize the $P_{50}$ quantile q-error within that group:

\begin{equation}
P_{50}\left(\left\{ \frac{\max(y_{ij}, P_i)}{\min(y_{ij}, P_i)} \mid j = 1, \ldots, n_i \right\}\right)
\end{equation}

The final overall approximate lower bound can be generated after calculating each \(P_i\). This lower bound applies to both neural network-based models and the XGBoost model. However, the bound is tighter for neural network-based models relative to the XGBoost model, since the XGBoost encoding is more lossy as it does not take
%considering the more lossy encoding of the XGBoost model without taking the plan
structural information into consideration. 

\begin{figure}[h]
    \centering
    \begin{minipage}{0.48\linewidth}
        \centering
        \includegraphics[width=\linewidth]{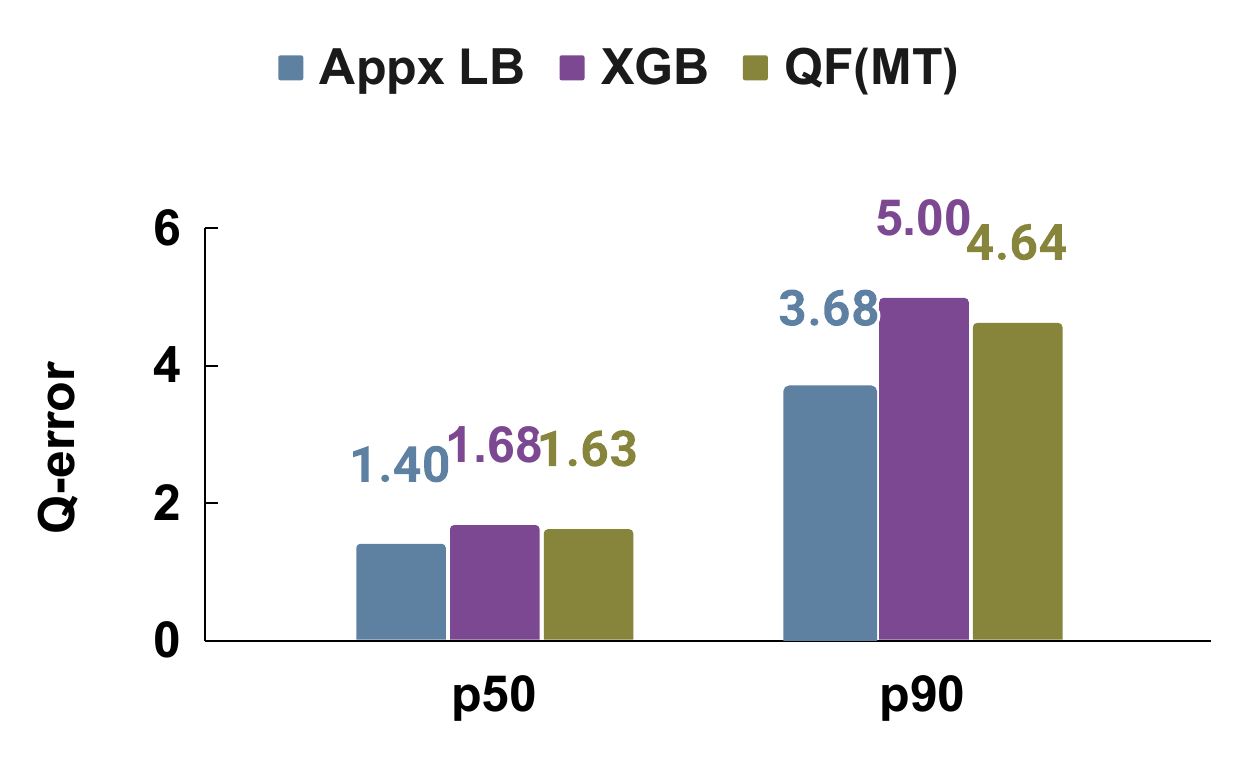}
        \caption{Comparison with empirical lower bound for latency prediction}
        \label{fig:lower_accuracy}
        \Description{Number of Spark queries per day across a month}
    \end{minipage}\hfill
    \begin{minipage}{0.48\linewidth}
        \centering
        \includegraphics[width=\linewidth]{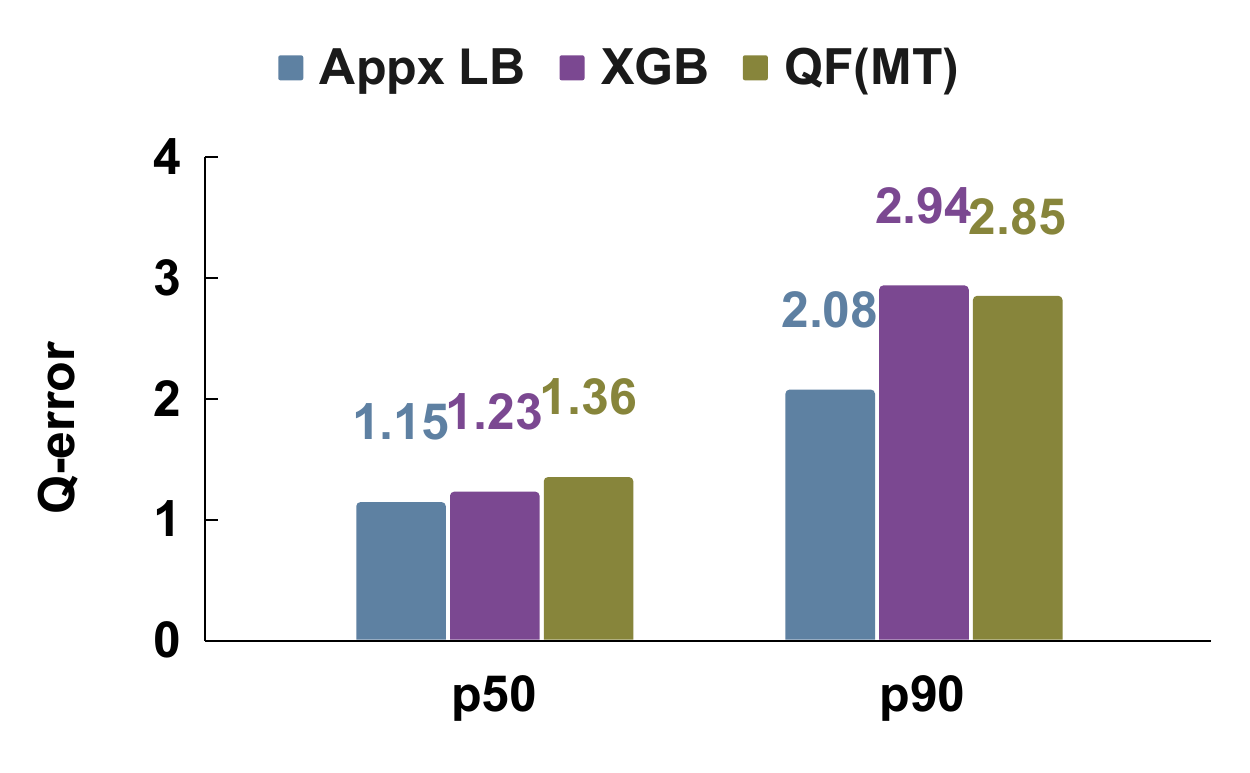}
        \caption{Comparison with empirical lower bound for CPU time prediction}
        \label{fig:lower_cpu}
        \Description{Number of Spark queries per day across a month}
    \end{minipage}
\end{figure}

%The resultant accuracies from this method and the comparison with the 
The comparison of these lower bound calculations with the results from QueryFormer(MT) and XGBoost are shown in Figures \ref{fig:lower_accuracy} and \ref{fig:lower_cpu}. 
%This calculated approach offers an empirical estimation of the maximum achievable model prediction accuracy. 
The figures illustrate that the prediction accuracy of both QueryFormer(MT) and XGBoost models is close to the approximate theoretical optimum. This result helps to explain the ceiling effect in model accuracy, in which accuracy did not significantly improve when training on a larger dataset in Section \ref{sec:largerdataset}.
%when we try to train on a larger dataset.  

% This observation empirically suggests why a comparatively simpler XGBoost model can achieve performance on par with more complex neural network models; it appears both types of models have approached the appropriate upper accuracy bound, 

% \begin{figure*}
%     \begin{subfigure}[b]{0.48\linewidth}
%         \includegraphics[width=\linewidth]{figures/lower_latency.png}
%         \caption{Comparison with lower bound for latency prediction}
%         \label{fig:lower_accuracy}
%         \Description{Number of Spark queries per day across a month}
%     \end{subfigure}
%     \hfill
%     \begin{subfigure}[b]{0.48\linewidth}
%         \includegraphics[width=\linewidth]{figures/lower_cpu.png}
%         \caption{Comparison with lower bound for CPU time prediction}
%         \label{fig:lower_cpu}
%         \Description{Number of Spark queries per day across a month}
%     \end{subfigure}
% \end{figure*}

This observation also helps to explain why reducing the number of classes leads to improved classification outcomes, as discussed in Section \ref{sec:num_class}. With a smaller number of classes, there is a higher likelihood of assigning consistent labels to query plans. 
%Consider the scenario where two plans share similar structures yet differ in their performance metrics. 
With a coarse-grained classification scheme, it is more likely that two plans that share similar structures yet differ in their performance metrics will be categorized into the same class.
%becomes more probable, ensuring uniform labels. 
In contrast, with a fine-grained classification scheme, these similar plans might be assigned disparate labels, which worses the theoretical lower bound on accuracy.
%, introducing challenges for the model. Thus, a reduction in 
Reducing the number of classes enhances label consistency, which in turn boosts classification accuracy. 

\section{Related Work}
\label{sec:related}
\noindent \textbf{Machine Learning-Based Query-Driven Approaches:}
Ganapathi et al.~\cite{DBLP:conf/sigmod/KimJSHCC22} investigate the application of Kernel Canonical Correlation Analysis (KCCA)~\cite{bach2002kernel} to predict the runtime and resource usage characteristics of queries for data warehouse management. Akdere et al.~\cite{DBLP:conf/icde/AkdereCRUZ12} analyze the trade-offs between making predictions at the plan-level and operator-level for query encodings and propose a hybrid encoding scheme that combines the benefits of both. The models they choose include Support Vector Machines (SVM)~\cite{cortes1995support}. Ding et al.~\cite{DBLP:conf/sigmod/DingDM0CN19} advocate for a classification approach over regression methods to improve the accuracy of index tuners. The models they explore include Logistic Regression (LR), Random Forest (RF), Gradient-Boosted Trees (GBT), and Light GBM with gradient-boosted decision trees (LGBM). Saxena et al.~\cite{DBLP:conf/sigmod/SaxenaRCLCCMKPN23} employ XGBoost~\cite{DBLP:conf/kdd/ChenG16}, leveraging optimizer-generated statistics to forecast the performance and memory requirements of queries within the cloud-based data warehouse Amazon Redshift~\cite{DBLP:conf/sigmod/GuptaATKPSS15}.

\noindent \textbf{Deep Learning-Based Query-Driven Approaches:}
QPPNet~\cite{DBLP:journals/pvldb/MarcusP19} employs compact neural units to emulate logical operators, constructing a network whose structure is isomorphic with that of the query plan. TLSTM~\cite{DBLP:journals/pvldb/SunL19} utilizes a Tree LSTM~\cite{DBLP:conf/acl/TaiSM15} to encapsulate a query plan's structural information.  QueryFormer~\cite{DBLP:journals/pvldb/ZhaoCSM22} adapts the Transformer architecture~\cite{DBLP:journals/corr/VaswaniSPUJGKP17} to represent the tree structure of the query plan.  Prestroid~\cite{DBLP:conf/sigmod/Kang0TCSH21} introduces a simplified tree convolution-based model to estimate Presto's resource consumption, prioritizing a reduced memory footprint and faster training time.

\noindent\textbf{Data-Driven Approaches:}
DeepDB~\cite{DBLP:journals/pvldb/HilprechtSKMKB20} introduces Relational Sum Product Networks (RSPNs), a variation of Sum Product Networks~\cite{DBLP:conf/iccvw/PoonD11}, to compute joint distributions. Naru~\cite{DBLP:journals/pvldb/YangLKWDCAHKS19} employs autoregressive models, which have demonstrated success in learning from images~\cite{DBLP:conf/iclr/SalimansK0K17}, audio~\cite{DBLP:conf/ssw/OordDZSVGKSK16}, and text~\cite{DBLP:journals/corr/VaswaniSPUJGKP17}, to address the cardinality estimation challenge on a single table. Furthermore, Naru introduces a novel Monte Carlo integration technique, termed progressive sampling, to efficiently estimate range queries. Without this integration technique, this task is computationally intensive when using a naive application of autoregressive models. Building upon this, NeuroCard~\cite{DBLP:journals/pvldb/YangKLLDCS20} expands Naru's capabilities to accommodate join operations. 
Face~\cite{DBLP:journals/pvldb/WangCLL21} capitalizes on a normalization flow (NF)~\cite{DBLP:journals/pami/KobyzevPB21} model and proposes dequantization techniques that reconcile the continuous attributes of NF models with discrete data.

\noindent \textbf{Learned Cardinality Estimation Methods Evaluation:}
Ortiz et al.~\cite{DBLP:journals/corr/abs-1905-06425} implement and evaluate a variety of deep learning models, such as fully connected neural networks and Recurrent Neural Networks (RNN), for cardinality estimation. They benchmark these models against tree ensemble methods including Random Forest and AdaBoost\cite{freund1997decision}, as well as against the PostgreSQL Optimizer. These models are compared on a variety of datasets, including IMDB~\cite{DBLP:conf/cidr/LeisRGK017}, DMV~\cite{DBLP:journals/pvldb/KieferHBM17}, and TPC-H. However, these evaluations do not incorporate newly proposed deep learning models that are specifically designed for cardinality estimation (CE). Our study differentiates from this prior work by including these models.
%, which we test. 
Furthermore, the tested workloads may not be sufficiently complex or representative enough of real-world queries, whereas our study is performed on real-world LinkedIn production workloads. Wang et al.~\cite{DBLP:journals/pvldb/WangQWWZ21} also compare the performance of various query-driven models—including MSCN~\cite{DBLP:conf/cidr/KipfKRLBK19}, LW-NN~\cite{DBLP:journals/pvldb/DuttWNKNC19}, LW-XGB~\cite{DBLP:journals/pvldb/DuttWNKNC19}, and DQM-Q~\cite{DBLP:conf/sigmod/HasanTAK020}—and data-driven models such as Naru~\cite{DBLP:journals/pvldb/YangLKWDCAHKS19} and DeepDB~\cite{DBLP:journals/pvldb/HilprechtSKMKB20}, against traditional cost models in both static and dynamic environments characterized by frequent data updates. 
Kim et al.~\cite{DBLP:conf/sigmod/KimJSHCC22} evaluate several query-driven and data-driven learned models; Similar to the work by Ortiz et al. discussed above, both the works by Wang et al. and Kim et al. perform evaluations on datasets that may not be sufficiently representative of real-world workloads, whereas our study is performed on real-world LinkedIn production workloads. Moreover, Wang et al do not assess newer models tailored for tree-structured query plans. %have yet to be assessed. 

\section{Conclusion}
\label{sec:conclusion}

This paper evaluated several learned query performance prediction models (TLSTM, TCNN, QueryFormer, and XGBoost) against real-world industrial workloads at LinkedIn. This analysis extended beyond traditional latency regression tasks to include CPU time prediction and classification tasks. For most tasks, we found that XGBoost performed the best of these existing approaches, both in terms of training time and accuracy. We also found that 
%introduce a 
incorporating a multi-task learning approach designed to enhance model capabilities can enable a deeper understanding of query plan performance. When integrating it with QueryFormer, we found that it achieves similar accuracy to XGBoost, and often outperforms it.

%Despite the inherent complexity and interpretative challenges associated with neural networks, we endeavor to quantitatively estimate the theoretical upper bounds of model accuracy from a data perspective. This approach aids in identifying the limitations imposed by query encoding and provides insights into potential avenues for refining predictive accuracy in complex industrial settings.

%\clearpage

\bibliographystyle{ACM-Reference-Format}
\bibliography{ref}

\end{document}